\documentclass[eqsecnum,floats,aps,superscriptaddress, nofootinbib,prd,showpacs]{revtex4}
\usepackage[T1]{fontenc}
\usepackage[utf8]{inputenc}
\usepackage{lmodern}
\usepackage{amsmath,amssymb,amsfonts}
\usepackage{hyperref}
\usepackage[english]{babel}
\usepackage{graphicx}
\usepackage{color}
\def\be{\nopagebreak[3]\begin{equation}}
\def\ee{\end{equation}}
\def\ba{\nopagebreak[3]\begin{eqnarray}}
\def\ea{\end{eqnarray}}
\def\bas{\nopagebreak[3]\begin{eqnarray*}}
\def\eas{\end{eqnarray*}}

\def\d{{\rm d}}
\def\a{\alpha}

\def\S{\Sigma}

\def\g{\gamma}

\def\ed{\d \!\!\!\! \d\, }

\newcommand{\f}{\frac}

\usepackage{enumerate}

\usepackage{colordvi}
\newcounter{mnotecount}[section]

\def\de{\delta}
\def\o{\omega}

\def\O{\Omega}
\def\D{\Delta}

\def\ula{\underleftarrow}
\def\heq{\hat{=}}
\def\bm{\bar{m}}
\def\eps{{}^2\!\epsilon}
\def\w{\wedge}

\def\Lb{\mathbf{L}}

\def\f{\frac}

\def\La{\pounds}
\def\a{\alpha}

\def\k{\kappa}

\newcommand{\re}{\mathbb{R}}

\def\Po{\mathrm{Po}}
\def\Eu{\mathrm{E}}
\def\ny{\mathrm{NY}}

\begin{document}

\title{Hamiltonian and Noether charges in first order gravity}
\author{Alejandro Corichi}\email{corichi@matmor.unam.mx}
\affiliation{Centro de Ciencias Matem\'aticas, Universidad Nacional Aut\'onoma de
M\'exico, UNAM-Campus Morelia, A. Postal 61-3, Morelia, Michoac\'an 58090,
Mexico}
\affiliation{Center for Fundamental Theory, Institute for Gravitation and the Cosmos,
Pennsylvania State University, University Park
PA 16802, USA}
\author{Irais Rubalcava-Garc\'\i a}
\email{irais@matmor.unam.mx}
\affiliation{Instituto de F\'{\i}sica y
Matem\'aticas,  Universidad Michoacana de San Nicol\'as de
Hidalgo, Morelia, Michoac\'an, Mexico}
\affiliation{Centro de Ciencias Matem\'aticas, Universidad Nacional Aut\'onoma de
M\'exico, UNAM-Campus Morelia, A. Postal 61-3, Morelia, Michoac\'an 58090,
Mexico}
\author{Tatjana Vuka\v sinac}
\email{tatjana@umich.mx}
\affiliation{Facultad de Ingenier\'\i a Civil, Universidad Michoacana de San Nicol\'as de
Hidalgo,\\ Morelia, Michoac\'an 58000, Mexico}
\begin{abstract}
We consider gravity in four dimensions in the vielbein formulation,  where the fundamental variables are
a tetrad $e$ and a SO(3,1) connection $\omega$. We start with the most general action principle compatible with diffeomorphism invariance which includes,
besides the standard Palatini term, other terms that either do not change the equations of motion, or 
are topological in nature. For our analysis we employ the covariant 
Hamiltonian formalism where the phase space $\Gamma$ is given by solutions to the equations of motion. 
We consider spacetimes that include a boundary at 
infinity, satisfying asymptotically flat boundary conditions and/or an internal boundary satisfying isolated horizons boundary conditions. 
For this extended action we study the effect of the topological terms on the Hamiltonian formulation.
We prove two results. The first one is rather generic, applicable to any field theory with boundaries: 
The addition of topological terms (and any other boundary term) does not modify the symplectic structure of the theory. 
The second result pertains to the conserved
Hamiltonian and Noether charges, whose properties we analyze in detail, including their relationship.
While the Hamiltonian charges are unaffected by the addition of topological and boundary terms, we show in detail that the Noether charges {\em do} change. Thus, a non-trivial 
relation between these two sets of charges arises when the boundary and topological terms needed for a consistent formulation are included.

\end{abstract}

\maketitle


\section{Introduction}


One of the main lessons from the general theory of relativity is that the theory is diffeomorphism invariant. It was the first example of such theories, where the theory is 
formulated without  background fields and the geometry of the underlying
background manifold becomes dynamical together with the rest of the fields present. The fact that the symmetry group of 
such theories corresponds to  diffeomorphisms on spacetime, poses new challenges for our understanding of the interplay between gauge and symmetries, and the existence of the 
corresponding conserved quantities. 
It is then of utter importance to have a thorough understanding of these issues starting from the
most basic principles. Namely, one requires a well defined action principle as the starting point for studying this kind of theories.
The importance of having such a requirement stems from the fact that  the classical theory is only a (very useful indeed!) approximation to a deeper underlying theory that must be 
quantum in nature. If, for 
instance, we think of a quantum theory defined by some path integral, in order for this to be well defined, we need to be able to write a meaningful action for the whole space of 
histories, and not only for the space of classical solutions. This observation becomes particularly vexing when the physical situation under consideration involves a spacetime 
region with boundaries. One must 
be particularly careful to extend the formalism in order to incorporate such boundary terms.

Yet another important issue in the definition of a physical theory is that of the choice of fundamental variables, specially when gauge symmetries are present. Again, even 
when the space of
solutions might coincide for two formulations, the corresponding actions will in general be different and that will certainly have an effect in the path integral formulation of 
the quantum theory. In the case of general relativity, that is the subject of this article, the better known formulation is of course in terms of a metric tensor $g_{ab}$, 
satisfying second order (Einstein) equations.
But there are other choices of variables that yield alternative descriptions. Here we shall consider one of those possibilities. In particular, the choice we shall make is 
motivated by writing the theory as a {\it local} gauge theory under the Lorentz group.
It is well known that one can either obtain Einstein equations of motion by means of the Einstein-Hilbert action, or in terms of the {\em Palatini action}, a first order action 
in terms of tetrads $e_a^I$ and a connection $\omega_a^{IJ}$ valued on the Lie Algebra of $SO(3,1)$ 
(see, e.g. \cite{romano} and \cite{peldan})\footnote{One should recall that the 
original Palatini action was written in terms of the metric $g_{ab}$ and an affine connection ${\Gamma^a}_{bc}$ \cite{palatini,adm}. The action we are considering here, in the so 
called ``vielbein" formalism, was developed in \cite{utiyama,kibble,sciama} and in \cite{di} in the canonical formulation.}.
Also, it is known that we can have a generalization of this action by adding a term, the Holst term, that still gives the same equations of motion and also allows to express 
the theory in terms of real $SU(2)$ connections in its canonical description (see, e.g. \cite{holst} 
 and \cite{barros}). This action, known as the Holst action, is the starting point for loop quantum gravity and some spin foam models. 
In the same first order scheme one can look 
for the most general diffeomorphism invariant first order action that classically describes general relativity, which can be written as the Palatini action (including the Holst 
term) plus topological contributions, namely, the Pontryagin, Euler and Nieh-Yan terms (see for instance
\cite{topo} for early references). Furthermore, if the spacetime region we are considering possesses boundaries one might have 
to add extra terms (apart from the topological terms that can also be seen as boundary terms) to the action principle\footnote{We should clarify our use of the name `topological term'. For us a term is topological if it can be written as a total derivative. This in turn implies that it does not contribute to the equations of motion. There are other possible terms that do not contribute to the equations of motion but that can not be written as a total derivative (such as the so called Holst term). For us, this term is not topological.}.

Thus, the most general first order action for gravity has the form,
\begin{equation}\label{1}
S[e, \omega] = S_{\rm Palatini} + S_{\rm Holst Term} + S_{\rm Pontryagin} + S_{\rm Euler} + S_{\rm Nieh-Yan} + S_{\rm Boundary}.
\end{equation}
It is noteworthy to emphasize that in the standard treatment of Hamiltonian systems one usually considers compact spaces without boundary, so there is no need to worry 
about the boundary terms that come from the integration by parts in the variational principle. But if 
one is interested in spacetimes with boundaries we can no longer neglect these boundary terms and it is mandatory to analyze them carefully. 
In order to properly study this action in the whole spacetime with boundaries, we need the action principle to be well posed, i.e. we want the action to be {\em differentiable 
and finite} under the appropriate boundary conditions, and under the most general variations compatible with the boundary conditions.

It has been shown that under appropriate boundary conditions, the Palatini action plus a boundary term provides a well posed action principle, that is, it is differentiable and 
finite.\footnote{See e.g. \cite{aes}, \cite{afk}, \cite{apv} and references therein for the asymptotically flat, isolated 
horizons and asymptotically AdS spacetimes respectively.
} Furthermore, in \cite{cwe} the analysis for 
asymptotically flat boundary conditions was extended to include the Holst term. Here we will refer to this well posed Holst action as the generalized Holst action (GHA). 

In order to explore some properties of the theories defined by an action principle, the covariant Hamiltonian formalism seems to be particularly appropriate (see, e.g. \cite{abr}, 
\cite{witten} and \cite{wald-lee}). In this formalism, one can introduce the standard Hamiltonian structures such as a phase space, symplectic structure, canonical transformations 
etc, without the need of a $3+1$ decomposition of the theory. All the relevant objects are {\it covariant}. The most attractive feature of this formalism is that one can find all 
these structures
in a direct fashion given the action principle. Furthermore one can, in a `canonical' way, find conserved quantities. On  the one hand one can derive Hamiltonian generators of 
canonical transformations and, on the other hand, Noetherian conserved quantities associated to symmetries, as e.g. in 
\cite{iw}. One important and interesting issue is to understand the precise 
relation between these two sets of quantities. 

The study of field theories with boundaries in the Hamiltonian approach has received certain attention in the literature. Most of these studies have focused on the standard formalism where a decomposition is involved and constraints are present. One recent example is \cite{barbero1}, that
considers linear gauge systems in the presence of boundaries, both in the Hamiltonian and covariant Hamiltonian frameworks, with an emphasis on the geometric approach and the functional analytic aspects of the problem (see the references there for previous studies). However, a detail study of a diffeomorphism invariant theory from this perspective is, in our opinion, still lacking.

The purpose of this article is to explore some of these issues in a systematic way. 
For that we shall study the action (\ref{1}), that has been shown to be well posed  for two sets of boundary conditions that are physically interesting; as outer boundary we shall 
consider 
configurations that are asymptotically flat, and in an inner boundary, those histories that satisfy isolated horizon boundary conditions \cite{crv1}. More concretely, we have two 
main goals.  The first objective is to explore the 
most basic structures in the covariant phase space formulation, where the phase space is given by
solutions to the equations of motion. More precisely, we shall study the dependence of the symplectic structure on the various topological and boundary 
terms in the action. As we show in detail, for any well posed field theory, the conserved symplectic structure is unaffected by  the addition of topological and other boundary terms. This 
result, however simple, seems to be repeatedly overlooked in the literature (one should note though, that  such statements are sometimes found within the context of standard canonical $3+1$ methods).
 The second goal is to explore the different conserved quantities that can be defined. Concretely, we  consider Hamiltonian 
conserved charges both at infinity and at the horizon. Finally, we  construct the associated Noetherian conserved current and charges. 
In both cases we shall study in detail how these quantities depend on the existence of the boundary terms that make the action well defined. As we shall show, while the 
Hamiltonian charges are insensitive to those quantities --given that the symplectic structure is invariant--, the Noether charges {\it do} depend on the form of the boundary terms 
added. 
Of particular interest is the notion of energy, as the quantity associated to asymptotic time 
translations. From the Hamiltonian point of view, it corresponds to the quantity generating those translation on phase space, while the Noetherian quantity is the conserved 
quantity associated to the invariance of the theory under such time translations. The fact that there are instances in which these quantities do not coincide is indeed puzzling. 

The structure of the paper is as follows:
In Section~\ref{sec:2} we present a brief review of the covariant Hamiltonian formalism,  in the cases when the spacetime has 
boundaries. We begin by defining the covariant phase space, and introduce the symplectic structure with its 
ambiguities and its dependence on boundary terms in the action. Finally we define the symplectic current structure, and the Hamiltonian and Noether charges. Here we prove our 
first result.
In Section~\ref{sec:3} we recall the most general action for general relativity in the first order formalism, as studied in \cite{crv1}. 
In particular, we consider spacetimes with boundaries: Asymptotically flatness at 
the outer boundaries, and an isolated horizon as an internal one.
In Section~\ref{sec:4} we study symmetries and their generators for both sets of boundary conditions. In particular we first compute the Hamiltonian conserved charges, and in the 
second part, the corresponding Noetherian quantities are found. We comment on the difference between  them. We summarize and provide some discussion in the final Section~\ref{sec:5}.


\section{Covariant Hamiltonian Formalism and conserved charges}
\label{sec:2}

In this section we give a self-contained review of the {\em covariant Hamiltonian formalism} (CHF) 
taking special care of the cases where boundaries are present.
It contains three parts. In the first one, we introduce the relevant structure in the definition of the 
covariant phase space, starting from the action principle. In particular, we see that boundary terms that 
appear in the `variation' of the action are of particular relevance to the construction of the symplectic 
structure. We shall pay special attention to the presence of boundary terms in the original action and how 
that gets reflected in the Hamiltonian formulation.  We prove the first result of this article.
In the second part, we recall the issue of symmetries 
of the theory. That is, when there are certain symmetries of the underlying spacetime, these get reflected 
in the Hamiltonian formalism. Of particular relevance is the construction of the corresponding conserved 
quantities, that are both conserved and play an important role of being the generators of such symmetries. 
In particular we focus our attention on the symmetries generated by certain vector fields, closely related 
to the issue of diffeomorphism invariance. In the third part we compare and contrast these Hamiltonian 
conserved quantities with the so-called Noether symmetries and charges. We show how they are related and 
comment on the fact that, contrary to the Hamiltonian charges, the corresponding `Noetherian' quantities 
{\em do} depend on the existence of boundary terms in the original action.

\subsection{Covariant Phase Space}
\label{sec:2.1}


In this part we shall introduce the relevant objects that define the covariant phase space.
If the theory under study has a \emph{well posed initial value formulation},
then, given the initial data we have a unique solution to the equations of motion. In this way we have 
an isomorphism ${I}$ between the space of solutions to the equations of motion, $\Gamma$, and the space of 
all valid initial data, the `canonical phase space' $\tilde{\Gamma}$. In this even dimensional space
 we can construct a nondegenerate, closed $2-$form $\tilde{\Omega}$, the symplectic form. Together, the phase space 
and the symplectic form constitute a symplectic manifold $(\tilde\Gamma,\tilde\Omega)$.

We can bring the symplectic structure to the space of solutions, via the pullback ${I}^*$ of $\tilde{\Omega}$ and 
define a corresponding 2-form on $\Gamma$.
In this way the space of solutions is equipped with a \emph{natural}
symplectic form, $\Omega$, since the mapping is independent of the reference Cauchy surface one is using to 
define ${I}$. Together, the space of solutions and its symplectic structure $(\Gamma, \Omega)$ are known
as the \emph{covariant phase space} (CPS).

However, most of the field theories of interest present \emph{gauge} symmetries. This fact is reflected on 
the symplectic form $\Omega$, making it \emph{degenerate}. When this is the case, $\Omega$ is only a pre-symplectic 
form, to emphasize the degeneracy. It is only after one gets rid of this degeneracy, by means of an appropriate 
quotient, that one recovers a physical non-degenerate symplectic structure. Let us now see how one can
arrive to such description from the action principle.

Before proceeding we shall make some remarks regarding notation. It has proved to be useful to use differential forms to deal with certain diffeomorphism invariant theories, and 
we shall do that here. However,
when working with differential forms in field theories one has to distinguish between the exterior derivative $\ed$ in the infinite dimensional covariant phase space,  and the 
`standard' exterior derivative on the spacetime manifold, denoted by $\d$. 
In this context, differential forms in the CPS act on vectors tangent to the space of solutions $\Gamma$. We use $\delta$ or $\delta \phi$ to denote tangent vectors, to be 
consistent with the standard notation used in  the literature.   We hope that no confusion should arise by such a choice. Let us now recall some basic constructions on the 
covariant phase space.

Taking as starting point an action principle, 
\begin{equation}
S[\phi^{A}] = \int_{\mathcal{M}}  \mathbf{L}\, ,\label{original_action}
\end{equation}
where the Lagrangian density, $\mathbf{L}$, is a $4-$form, that depends on fields $\phi^A$ and their derivatives.
The fields $\phi^{A}$ are certain $n-$forms (with $n \leq 4$) 
in the $4-$dimensional spacetime manifold, $\mathcal{M}$, with boundary, $\partial\mathcal{M}$, and $A,B,\ldots$ are internal indices.
Then, the variation of the action can be written as,
\begin{equation}\label{VarActFormsWithoutBoundary}
\ed S (\delta)  = \delta S = \int_{\mathcal{M}} E_{A} \wedge \delta \phi^{A} + \int_{\mathcal{M}} \d \theta ( \delta \phi^{A})\, ,
\end{equation}
where $E_{A}$ are the Euler-Lagrange equations of motion forms
and $\delta \phi^{A}$ is an arbitrary vector, that can be thought 
to point `in the direction that $\phi^{A}$ changes'. The 1-form (in CPS) $\theta$ depends on $\phi^A$, $\delta\phi^A$ and their derivatives, for
simplicity we do not write it explicitly. Note that we are using $\delta \phi^{A}$ and $\delta$, to denote \emph{the same} 
object. For simplicity in the notation, sometimes the $\phi^{A}$ part is dropped out. Here we 
wrote both for clarity. 
The second term of the RHS is obtained after integration by parts, and using Stokes' theorem it can be written as,
\begin{equation}\label{SympPotentialWhitoutBoundary}
\Theta (\delta \phi^{A}) := \int_{\mathcal{M}} \d \theta ( \delta \phi^{A}) = \int_{\partial \mathcal{M}} \theta ( \delta \phi^{A})\, .
\end{equation}
This term can be seen as a $1-$form in the covariant phase space, acting on vectors $\delta \phi^{A}$ and returning a real number. Also it can be seen as a potential for the 
symplectic structure, that we already mentioned in the preamble of this section and shall define below. For such a reason, we will call this term, 
$\Theta(\delta \phi^{A})$  a 
\emph{symplectic potential} associated to a boundary $\partial\mathcal{M}$, and the integrand, $\theta (\delta \phi^{A})$, is the 
\emph{symplectic potential current}\footnote{Usually, a symplectic potential is defined as an integral of $\theta$ over a spatial slice $M$, see, for example,
\cite{abr}. Here, we are extending this definition since, as we shall show, in order to construct a symplectic structure it is important to consider the integral
over the whole boundary $\partial\mathcal{M}$.}.

Note that from Eqs. (\ref{VarActFormsWithoutBoundary}) and (\ref{SympPotentialWhitoutBoundary}), in the space of solutions $E_{A} = 0$, 
$\ed S = \Theta (\delta \phi^{A})$.

If we want to derive in a consistent way the equations of motion for the system, 
the action must be differentiable. In particular, this means that we need the boundary term 
(\ref{SympPotentialWhitoutBoundary}) to be zero. To simply demand that  
$\delta \phi^{A} |_{\partial \mathcal{M}} = 0$, becomes too restrictive if we want to allow {\em all}  
the variations which preserve appropriate boundary conditions and not just variations 
of compact support. Thus, requiring the action to be stationary with respect to all compatible variations 
should yield precisely the classical equations of motion, with the respective boundary term vanishing 
on any allowed variation.

If the original action is not well defined, the introduction of a boundary term could be needed. In that case the action becomes,
\begin{equation}\label{ActionFormsWithBoundary}
S[\phi^{A}] = \int_{\mathcal{M}}  \mathbf{L} + \int_{\mathcal{M}} \d \varphi \, ,
\end{equation}
where the boundary term in general depends on fields, as well as of their derivatives, 
and is chosen in such a way that the new action is differentiable and finite, for allowed field configurations, 
and we have a well posed variational principle,
\begin{equation}\label{GeneralVariatonForms}
\delta S = \int_{\mathcal{M}} E_{A} \wedge \delta \phi^{A} + \int_{\mathcal{M}} \d \left[\theta ( \delta \phi^{A}) +  \delta  \varphi  \right].
\end{equation}
When we have added a boundary term, the symplectic potential associated to this well posed action changes as $\Theta \rightarrow \Theta  + \int_{\mathcal{M}} \d \delta \varphi$, equivalently we can consider,
\begin{equation}
\tilde \Theta (\delta ) := 
\int_{\partial \mathcal{M}}  \left[\theta (\delta ) +  \delta  \varphi  \right] .\label{sympl_poten_boundary_term}
\end{equation}
From this equation we can see that besides the boundary term added to the action, to make it well defined, we can always add a term, $\d Y$, to the symplectic potential current 
that will not change $\tilde{\Theta}$. Thus, the most general symplectic potential can be written as,
\begin{equation}\label{GeneralTheta}
\tilde{\Theta} ( \delta ) = 
\int_{\partial \mathcal{M}}  \left[\theta ( \delta ) +  \delta  \varphi + 
\d Y( \delta ) \right] =: \int_{\partial \mathcal{M}}  
\tilde{\theta} ( \delta )\, .
\end{equation}


Now, we take the exterior derivative of the symplectic potential, $\tilde{\Theta}$, acting on tangent vectors $\delta_{1}$ and $\delta_{2}$ at a point $\gamma$ 
of the phase space,
\begin{equation}\label{edTheta}
\ed \tilde{\Theta} (\delta_{1}, \delta_{2}) = \delta_{1} \tilde{\Theta} (\delta_{2}) -\delta_{2} \tilde{\Theta} (\delta_{1}) 
= 2 \int_{\partial \mathcal{M}} \delta_{[1} \tilde{\theta} (\delta_{2]})\, .
\end{equation}
From this expression we can define a spacetime $3-$form, \emph{the symplectic current} $\tilde{J}(\delta_{1}, \delta_{2})$, to be the integrand of the RHS of (\ref{edTheta}),
\begin{equation}\label{DefJ}
\tilde{J}(\delta_{1}, \delta_{2}) := \delta_{1} \tilde{\theta}  (\delta_{2}) -\delta_{2} \tilde{\theta}  (\delta_{1})\, .
\end{equation}

In particular, when we have added a boundary term to the action, and taking into account the ambiguities, the symplectic current becomes,
\begin{equation}\label{defJ}
 \tilde{J}(\delta_{1}, \delta_{2}) = J(\delta_{1}, \delta_{2}) + 2\bigl(\delta_{[1} \delta_{2]} \varphi +  
 \delta_{[1}\d Y (\delta_{2]})\bigr)\, .
\end{equation}
where
\begin{equation}\label{defJ2}
J(\delta_{1}, \delta_{2}) := \delta_{1} \theta  (\delta_{2}) -\delta_{2} \theta  (\delta_{1})\, ,
\end{equation}
is the symplectic current associated to the action (\ref{original_action}).

Now, the term $ \delta_{[1} \delta_{2]} \varphi $ vanishes by antisymmetry, because $\delta_{1}$ and $\delta_{2}$ commute when acting on functions. Note that the last term of the 
RHS of (\ref{defJ}) can be written as $\d \chi (\delta_{1}, \delta_{2}) = 2\delta_{[1}\d Y (\delta_{2]})$ due to $\d$ and $\delta_{i}$ commuting. Since $\d$ and $\ed$ act on 
different spaces, the spacetime and the space of fields, respectively, they are independent. In this way $\tilde{J}(\delta_{1}, \delta_{2})$ is determined as
\begin{equation}\label{defJ1}
 \tilde{J}(\delta_{1}, \delta_{2}) = J(\delta_{1}, \delta_{2}) +  \d \chi (\delta_{1}, \delta_{2})\, .
\end{equation}
This ambiguity will appear explicitly in the examples that we shall consider below.

Therefore we conclude that, \emph{when we add a boundary term to the original action it will not change the symplectic current}, and this result holds independently of the 
specific boundary conditions. This is the first result of this article. 

Recall that in the space of solutions, $\ed S (\delta) = \tilde{\Theta} (\delta)$, therefore from eqs. (\ref{edTheta}) and (\ref{DefJ}),
\begin{equation}
0 = \ed ^{2} S (\delta_{1}, \delta_{2}) =  \ed \tilde{\Theta} (\delta_{1}, \delta_{2})   =  2 \int_{\mathcal{M}} \delta_{[1} \d 
\tilde{\theta} ( \delta_{2]} ) = \int_{\mathcal{M}} 
 \d\tilde{J}(\delta_{1}, \delta_{2}).
\end{equation}

\begin{figure}[h]
\begin{center}
  \includegraphics[width=8cm]{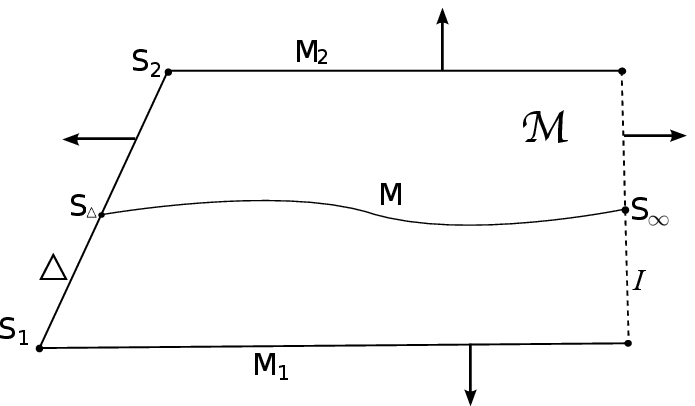}
  \caption{The region $\mathcal{M}$.}\label{regionM}
  \end{center}
\end{figure}

Since we are integrating over \emph{any} region $\mathcal{M}$, we can conclude that $\tilde{J}$ is closed, i.e. $\d\tilde{J} = 0$. Note that 
$ \d\tilde{J} =\d ( J +  \d \chi ) = \d J$ depends only on $\theta$. Using Stokes' theorem, and taking into account the orientation of $\partial\mathcal{M}$ 
(see Fig. \ref{regionM}), we have
\begin{equation}\label{intJzero}
0 = \int_{\mathcal{M}}  \d \tilde{J}(\delta_{1}, \delta_{2})   = \int_{\mathcal{M}}  \d J(\delta_{1}, \delta_{2}) =
 \oint_{\partial \mathcal{M}} J(\delta_{1}, \delta_{2})  =
\left(-  \int_{M_{1}} + \int_{M_{2}}  - \int_{\Delta}  + 
\int_{\mathcal{I}} \right)  J,
\end{equation}
where $\mathcal{M}$ is bounded by $\partial \mathcal{M} = M_{1} \cup M_{2} \cup  \Delta \cup \mathcal{I}$, $M_{1}$ and $M_{2}$ are space-like slices, $\Delta$ is an inner boundary and $\mathcal{I}$ an outer boundary.

Now consider the following two possible scenarios: First, consider the case when there is no internal boundary, only a boundary $\mathcal{I}$ at infinity. In some instances the 
asymptotic conditions ensure that the integral $\int_{\mathcal{I}}  J$  vanishes, in which case from (\ref{intJzero}), one gets
\begin{equation}
\oint_{\partial \mathcal{M}}   J(\delta_{1}, \delta_{2})  =\left( - \int_{M_{1}} + \int_{M_{2}}   \right)  J =0\, ,
\end{equation}
which implies that $\int_{M}  J$ is independent of the Cauchy surface. This allows us to define a \emph{conserved} pre-symplectic form over an arbitrary space-like surface $M$,
\begin{equation}\label{symplectic_structure}
\bar{\Omega} (\delta_{1}, \delta_{2}) = \int_{M}  J  (\delta_{1}, \delta_{2})\, .
\end{equation}
Note that in (\ref{intJzero}) at the end we only have a contribution from $J$, not from the complete $\tilde{J}$, and for that reason the pre-symplectic
form does not depend on $\varphi$ (the contribution of the topological, total derivative, terms in the action) nor  $\chi$ (the contribution
of total derivative terms in $\tilde{J}$).

One should have special care in the case when  
the symplectic current is of the form $J=J_0+\d\alpha$, as we shall now demonstrate. Our previous arguments, see (\ref{defJ1}) and (\ref{intJzero}), show 
that the $\d\alpha$ term does not appear in the symplectic structure. 
It follows then that, when $J_0=0$, the symplectic structure is trivial $\bar\Omega =0$, by construction, so that in the definition (\ref{symplectic_structure}), it is only the $J_0$ part of $J$ that contributes to $\bar\Omega$.
It should be obvious that this conclusion is valid also in the case when there is an internal boundary $\Delta$. We shall further comment on this case below.




Let us now consider with more details the case when we have an internal boundary. Now, the integral $ \int_{\Delta} J$ may no longer vanish under the boundary conditions, as is the case with the 
isolated horizon boundary conditions (more about this below). The ``next best thing'' is that this integral is ``controllable''. Let us be more specific. If, after imposing 
boundary conditions, the integral takes the form,
\begin{equation}\label{Jhorizon}
\int_{\Delta} J =  \int_{\Delta} \d j = \int_{\partial \Delta} j\, ,
\end{equation}
we can still define a \emph{conserved} pre-symplectic structure. From (\ref{intJzero}), and assuming the integral over the outer boundary vanishes, we now have
\begin{equation}
\left(-  \int_{M_{1}} + \int_{M_{2}}  - \int_{\Delta}  \right)  J = \left(-  \int_{M_{1}} + \int_{M_{2}}  \right)  J  
- \left(  \int_{S_{1}} - \int_{S_{2}} \right) j = 0\, ,
\end{equation}
where $S_{1}$ and $S_{2}$ are the intersections of space-like surfaces $M_{1}$ and $M_{2}$ with the inner boundary $\Delta$, respectively.
Therefore we can define the \emph{conserved} pre-symplectic structure as,
\begin{equation}\label{sympl_struct}
\bar{\Omega}  = \int_{M}  J   +  \int_{S} j\, .
\end{equation}
Note that by construction, the two form $\bar{\Omega}$ is closed, so it is justified to call it a (pre-)symplectic structure.

Let us end this section by further commenting on the case when the symplectic current contains a total derivative.  In the literature, the symplectic structure is sometimes defined, from the beginning, as an integral of $\tilde{J}$ over a spatial hypersurface $M$, but we have shown that this is correct only if $\tilde{J}$ does not have a total derivative term, and the action does not have a boundary term. 
Let us now describe the argument that one sometimes encounters in this context, in the simple case where $J=\d\alpha$. In this case one could {\it postulate} the existence of a pre-symplectic structure $\tilde\Omega_M$ as follows. Let us  define
\begin{equation}\label{symplectic_structure_spatial}
 \tilde\Omega_M (\delta_1,\delta_2) := \int_M \d\alpha (\delta_1,\delta_2)\, ,
\end{equation}
we have, from (\ref{intJzero}) that $\tilde\Omega_M$ is independent on $M$ only if $\int_{\mathcal{I}} \d\alpha$ 
{\it and} $\int_{\Delta}\d\alpha$ vanish. In this case the object $\tilde\Omega_M$ does define a conserved two-form that satisfies the definition of a pre-symplectic structure. It should be stressed though that
such an object does not follow from the systematic derivation we followed, starting from an action principle. It can instead be viewed and a possible freedom that exists in the covariant Hamiltonian formalism. It is indeed interesting to explore the possible physical consequences of introducing the object $\tilde\Omega_M$.
As we shall show in forthcoming sections, there is one instance in which one can postulate such a two-form, that satisfies the conditions for being conserved, but as we shall show in detail, one does run
into inconsistencies when postulating such object for topological terms.

To summarize, in this part we have developed in detail the covariant Hamiltonian formalism in the presence of boundaries. As we have seen, there might be a contribution to the 
(pre-)symplectic structure coming from the boundaries. Finally,
we have shown that the addition of boundary terms to the action does not modify the
conserved (pre-)symplectic structure of the theory, independently of the boundary conditions imposed.
This is the first result of this note. As a remark, we have also noted that under certain circumstances, one could introduce a conserved symplectic structure 
that results from the existence of a total derivative term in the symplectic current.

\subsection{Symmetries and conserved charges}
\label{sec:2.2}

Let us now explore how the covariant Hamiltonian formulation can deal with the existence of symmetries, and their associated conserved quantities. Before that, let us recall the 
standard notion of a Hamiltonian vector field (HVF) in Hamiltonian dynamics.
A Hamiltonian vector field $Z$ is defined as a symmetry of the symplectic structure, namely 
\begin{equation}
\La_{Z} \Omega = 0.
\end{equation}
From this condition and the fact that $\ed \Omega = 0$ we have,
\begin{equation}\label{LieOmegaZero}
\pounds_{Z} \Omega = Z\cdot\ed \Omega + \ed (Z\cdot \Omega) = \ed (Z\cdot \Omega) = 0.
\end{equation}
where $Z\cdot\Omega\equiv i_{Z} \Omega$ means the \emph{contraction} of the 2-form $\Omega$ with the vector field $Z$. Note that 
$(Z\cdot\Omega) (\delta) =  \Omega(Z, \delta)$ is 
a one-form on $\Gamma$ acting on an arbitrary vector $\delta$. We can denote it as $X (\delta): = \Omega (Z, \delta)$.
From the previous equation we can see that $X= Z\cdot \Omega$ is closed, $\ed X = 0 $. 
It follows from (\ref{LieOmegaZero}) and from the Poincar\'e lemma that locally (on the CPS), there exists a function $H$ such that $X = \ed H$.
We call this function, $H$, the \emph{Hamiltonian}, that generates the infinitesimal canonical transformation defined by $Z$. Furthermore, and by its own definition, $H$ 
is a \emph{conserved quantity} along the flow generated by $Z$.

Note that the directional derivative of the Hamiltonian $H$, along an arbitrary vector $\delta$ can be written in several ways,
\begin{equation}
X (\delta) = \ed H(\delta) = \delta H,
\end{equation}
some of which will be used in-distinctively in what follows.

So far this vector field $Z$ is an arbitrary Hamiltonian vector field on $\Gamma$. Later on we will relate it to certain space-time symmetries. For instance, for field theories 
that possess a symmetry group, such as the Poincar\'e group for field theories on Minkowski spacetime, there will be corresponding Hamiltonian vector fields associated to the 
generators of the symmetry group. In this article we are interested in exploring gravity theories that are diffeomorphism invariant. That is, such that the diffeomorphism group 
acts as a (kinematical) symmetry of the action. Of particular relevance is then to understand the role that these symmetries have in the Hamiltonian formulation. In particular, 
one expects that diffeomorphisms play the role of {\em gauge} symmetries of the theory. However, the precise form in which diffeomorphisms can be regarded as gauge or not, depends 
on the details of the theory, and is dictated by the properties of the corresponding Hamiltonian vector fields. Another important issue is 
to separate those diffeomorphisms that are gauge from those that represent truly physical canonical transformations that
change the system. Those true motions could then be associated to symmetries of the theory. For instance, in the case of asymptotically flat spacetimes, {\em some} diffeomorphism 
are regarded as gauge, while others represent nontrivial transformations at infinity and can be associated to the generators of the Poincar\'e group. 
In the case when the vector field $Z$ generates time evolution,  one expects $H$ to be related to the energy, the ADM energy at infinity. Other conserved
Hamiltonian charges can thus be found, and correspond to the generators of the asymptotic symmetries of the theory.

\subsection{Diffeomorphism invariance: Noether charge}
\label{sec:2.3}

Let us briefly review some results about the Noether current 3-form $J_N$ and its relation 
to the symplectic current $J$. For that, we shall rely on \cite{iw}. We know that to any Lagrangian theory 
invariant under diffeomorphisms we can associate 
a corresponding Noether current 3-form. Consider infinitesimal diffeomorphism generated by 
a vector field $\xi$. These diffeomorphisms
induce the infinitesimal change of fields, given by $\de_\xi\phi^A:=\La_\xi\phi^A$. 
From (\ref{VarActFormsWithoutBoundary}) it follows that the corresponding change in the 
lagrangian four-form is given by
\be
\La_\xi\Lb ={\rm E}_A\w\La_\xi\phi^A +\d\theta (\phi^A ,\La_\xi\phi^A )\, .\label{3.1}
\ee
On the other hand, using Cartan's formula, we obtain
\be
\La_\xi\Lb = \xi\cdot\d\Lb + \d (\xi\cdot\Lb )=\d (\xi\cdot\Lb )\, ,\label{3.2}
\ee
since $\d\Lb =0$, in a four-dimensional spacetime. Now, we
can define a {\it Noether current 3-form} as
\be
J_N (\de_\xi )=\theta (\de_\xi )-\xi\cdot\Lb\, ,\label{noether}
\ee
where we are using the simplified notation $\theta (\de_\xi ):=\theta (\phi^A ,\La_\xi\phi^A )$.
From the equations (\ref{3.1}) and (\ref{3.2}) it follows that
on the space of solutions, $\d J_N (\de_\xi )=0$, so at least locally one can define a
corresponding Noether charge density 2-form $Q_\xi$ relative to $\xi$ as 
\be
J_N (\de_\xi ) = \d Q_\xi\, .\label{nch}
\ee
Following \cite{iw}, the integral of $Q_\xi$ over some compact surface $S$ is the Noether
charge of $S$ relative to $\xi$.
As we saw in the previous chapter there are ambiguities in the definition of $\theta$
(\ref{GeneralTheta})  , that produce ambiguities in $Q_\xi$. 
As we saw in the section \ref{sec:2.1}, 
$\theta$ is defined up to an exact form: $\theta\to\theta +\d Y(\de )$. Also, the change in
Lagrangian $\Lb\to \Lb +\d\varphi$ produces the change $\theta\to\theta +\de\varphi$. 
These transformations affect the symplectic current in the following way
\be
J(\de_1,\de_2)\to J(\de_1,\de_2)+\d \bigl(\de_2 Y(\de_1)-\de_1 Y(\de_2)\bigr)\, .
\ee
The contribution of $\varphi$ vanishes, as before, and as we have shown in section \ref{sec:2.1}.
The above transformation leaves invariant the symplectic structure.
It is easy to see that the two changes, generated by $Y$ and $\varphi$ contribute to the 
following change of Noether current 3-form
\be\label{noether_current_change}
J_N(\de_\xi )\to J_N(\de_\xi )+\d Y(\de_\xi )+\de_\xi\varphi -\xi\cdot\d\varphi\, ,
\ee
and the corresponding Noether charge 2-form changes as
\be\label{noether_charge_change}
Q_\xi\to Q_\xi +Y(\de_\xi )+\xi\cdot\varphi +\d Z\, .
\ee
The last term in the previous expression is due to the ambiguity present in (\ref{nch}). This arbitrariness
in $Q_\xi$ was used in \cite{iw} to show that the Noether charge form of a general theory of gravity
arising from a diffeomorphism invariant Lagrangian, in the second order formalism, can be decomposed
in a particular way. 

Since $\d J_N (\de_\xi )=0$ it follows, as in (\ref{intJzero}), that 
\begin{equation}
 0 = \int_{\mathcal{M}}  \d J_N (\de_\xi )   = \oint_{\partial \mathcal{M}} J_N (\de_\xi )  
 = \left(-  \int_{M_{1}} + \int_{M_{2}}  - \int_{\Delta}  + \int_{\mathcal{I}} \right)J_N (\de_\xi ),
\end{equation}
and we see that if $\int_{\Delta}J_N (\de_\xi )=\int_{\mathcal{I}}J_N (\de_\xi )=0$ then the
previous expression implies the existence of the conserved quantity (independent on the choice
of $M$),
\begin{equation}
 \int_{M}J_N (\de_\xi )=\int_{\partial M}Q_\xi \, .
\end{equation}
Note that the above results are valid only on shell.

In the covariant phase space, and for $\xi$ arbitrary and fixed, we have \cite{iw}
\be
\de J_N (\de_\xi )=\de \theta (\de_\xi )-\xi\cdot \de\Lb =\de \theta (\de_\xi )-\xi\cdot\d\theta (\de )\, .
\ee
Since, $\xi\cdot\d\theta =\La_\xi \theta -\d (\xi\cdot\theta )$ and
$\de \theta (\de_\xi )-\La_\xi \theta (\de )=J(\de ,\de_\xi )$ by the definition of the 
symplectic current $J$ (\ref{DefJ}), it follows that the relation between the symplectic current $J$
and the Noether current 3-form $J_N$ is given by
\be
J(\de ,\de_\xi )=\de J_N (\de_\xi )-\d (\xi\cdot\theta (\de ))\, .\label{nc}
\ee
We shall use this relation in the following sections, for the various actions that describe
first order general relativity, to clarify the relation between the Hamiltonian and Noether charges.
We shall see that, in general, a Noether charge does not 
correspond to a Hamiltonian charge generating symmetries of the phase space.

\section{The action for gravity in the first order formalism}
\label{sec:3}

As already mentioned in the introduction, we shall consider the most general action for four-dimensional gravity in the first order formalism. 
The choice of basic variables is the following:
A pair of co-tetrads $e_a^I$ and a Lorentz $SO(3,1)$ connection $\o_{aIJ}$ on the spacetime $\mathcal{M}$, possibly with a boundary. 
In order for the action to be physically relevant,
it should reproduce the equations of motion for general relativity and
be: 1) differentiable, 2) finite on the configurations with a given asymptotic behaviour and
3) invariant under diffeomorphisms and local internal Lorentz transformations. The most general action that gives the desired
equations of motion and is compatible with the symmetries of the theory is given by the combination of
Palatini action, $S_{\mathrm{P}}$, Holst term, $S_{\mathrm{H}}$, and three topological terms, Pontryagin, 
$S_\Po$, Euler, $S_\Eu$, and Nieh-Yan, $S_\ny$, invariants. As we shall see, the Palatini term contains
the information of the ordinary Einstein-Hilbert 2nd order action, so it represents the backbone of
the formalism. Since we are considering a spacetime region $\mathcal{M}$ with boundaries, one should pay special attention to boundary conditions. For instance, it turns out that 
the Palatini action, as well as
Holst and Nieh-Yan terms are not differentiable for asymptotically flat spacetimes, and 
appropriate boundary terms should be provided. 

\subsection{The complete action}
\label{sec:3.1}

In \cite{crv1} we have shown that the ``most general'' first order 
diffeomorphism invariant action that classically describes general relativity, that is 
well defined in the case 
of asymptotically flat spacetimes with a weakly isolated horizon,  can be written as,
\begin{equation}\label{complete-action}
S[e, \omega] = S_{\rm PB} + S_{\rm H} + \alpha_{1} S_{\rm Po} + 
\alpha_{2}S_{\rm E} + \alpha_{3} S_{\rm NY} + \alpha_{4}S_{\rm BH}.
\end{equation}
Here $\alpha_{1}, ...,\alpha_{4}$ are coupling constants.
The coupling constants $\alpha_{1}$ and $\alpha_{2}$, are not fixed by our boundary conditions,
while different choices for the Holst-Nieh-Yan sector of the theory,  
imply particular combinations of $\alpha_3$ and $\alpha_4$. To see that, consider
$S_{\rm BH}$ that represents the boundary term that we need to add to Holst term in order to make it well defined. 
As shown in \cite{crv1}, if $\a_3=-\frac{1}{2\k\g}$ then the combination of the Holst
and Nieh-Yan terms is well defined and no additional boundary term is needed, so $\a_4=0$ in that
case. For every other value of $\a_3$ we need to add a boundary term, and in that case 
$\a_4=\frac{1}{2\k\g}+\a_3$.
Other than these cases, there is no important relation between the different coupling constants.


Palatini action with boundary term is given by \cite{aes},
\begin{equation}\label{PalatiniplusBoundary}
  S_{\mathrm{PB}} = - \frac{1}{2 \kappa} \int_{\mathcal{M}} \Sigma^{IJ} \wedge  F_{IJ}  +   
  \frac{1}{2 \kappa} \int _{\partial \mathcal{M}} \Sigma^{IJ} \wedge \omega_{IJ} .
\end{equation}
where $\k =8\pi G$, 
$\Sigma^{IJ} = \star (e^I\w e^J):=\frac{1}{2} \epsilon^{IJ}\,_{JK}e^{J}\wedge e^{K}$,
$F_{IJ}=\d\o_{IJ}+\o_{IK}\w{\o^K}_J$ is a curvature two-form of the connection $\o$
and, as before, $\partial \mathcal{M} = M_{1} \cup M_{2} \cup  \Delta \cup \mathcal{I}$.
The boundary term is not manifestly gauge invariant, but, as pointed out in \cite{aes}, it is effectively
gauge invariant on the spacelike surfaces $M_1$ and $M_2$ and also in the asymptotic region $\mathcal{I}$.
This is due to the fact that the only allowed gauge transformations that preserve the asymptotic conditions are such that the boundary terms remain invariant.

It turns out that at spatial infinity this boundary term does not reduce to Gibbons-Hawking surface term,
the later one is divergent for asymptotically flat spacetimes, as shown in \cite{aes}. Let us mention that
there have been other proposals for boundary terms for Palatini action, as for example in \cite{obukhov} and
\cite{bn}, that are equivalent to Gibbons-Hawking action and are obtained without imposing the time gauge condition. 
They are manifestly gauge invariant and well defined for finite boundaries, but they are not well defined
for asymptotically flat spacetimes. In time gauge they reduce to (\ref{PalatiniplusBoundary}).

The {\it Holst} term \cite{holst}, was first introduced with the aim of having a variational 
principle whose $3+1$ decomposition yielded general relativity in the  Ashtekar-Barbero (real) variables \cite{barbero}.
It turns out that the Holst term, when added to the Palatini action,
does not change the equations of motion
(although it is not a topological term), so that in the Hamiltonian formalism its addition corresponds to
a canonical transformation. This transformation leads to the Ashtekar-Barbero variables that are the basic
ingredients in the loop quantum gravity approach. It turns out that the Holst term is finite but not differentiable 
for asymptotically flat spacetimes, so an appropriate boundary term should be added in order 
to make it well defined. The result is \cite{cwe},
\begin{equation}\label{HolsttermBoundary}
  S _{\mathrm{HB}}= - \frac{1}{2 \kappa\g} \int_{\mathcal{M}} \Sigma^{IJ}\w \star F_{IJ}  +   
  \frac{1}{2 \kappa\g} \int _{\partial \mathcal{M}}  \Sigma^{IJ} \wedge \star\o_{IJ} \, ,
\end{equation}
where $\g$ is the Barbero-Immirzi parameter. 

In four dimensions there are three topological invariants constructed from $e^I$, $F_{IJ}$ and
$D e^I$, consistent with diffeomorphism and local Lorentz invariance. They are exact forms and
do not contribute to the equations of motion, but in order to be well defined they should be finite, 
and their variation  on the boundary of the spacetime region ${\mathcal{M}}$ 
should vanish. The first two terms, the Pontryagin and Euler terms are constructed from
the curvature $F_{IJ}$ and its dual (in the internal space) $\star F_{IJ}$, while the third one, 
the Neih-Yan invariant, is related to torsion $T^L:=De^L=\d e^L +{\o^L}_K \w e^K$.

 The action corresponding to the Pontryagin term is given by,
\be
\label{Pontryagin} S_{\Po} =\int_{\mathcal{M}} F^{IJ} \wedge F_{IJ} = 
2\int_{\partial \mathcal{M}} \left( \omega_{IJ} \wedge d\omega^{IJ} + 
\frac{2}{3} \omega_{IJ}\wedge \omega^{IK} \wedge \omega_{K}\,^{J} \right)\, .
\ee
The boundary term  is the Chern-Simons Lagrangian density, $L_{CS}\,$.

The action for the Euler term, is given by,
\be
\label{Euler} S_{\Eu} = \int_{\mathcal{M}} F^{IJ} \wedge \star F_{IJ}= 
2\int_{\partial \mathcal{M}} \left( \star \omega_{IJ} \wedge d\omega^{IJ} + 
\frac{2}{3} \star{\omega}_{IJ}\wedge \omega^{IK} \wedge \omega_{K}\,^{J} \right)\, .
\ee

The Nieh-Yan topological invariant is of a different nature from the two previous terms. 
It is related to torsion and its contribution to the action is \cite{nieh-yan,nieh},
\be
\label{Nieh-Yan} S_{\ny} = \int_{\mathcal{M}} \left( D e^{I} \wedge D e_{I}  - \Sigma^{IJ} \wedge \star F_{IJ} \right)= 
\int_{\partial \mathcal{M}} De^{I} \wedge e_{I}\, .
\ee
Note that the Nieh-Yan term can be written as
\begin{equation}\label{relationNYHolst}
S_{\ny} = 2\k\g S_{\mathrm{H}}+\int_{\mathcal{M}}  D e^{I} \w D e_{I}\, ,
\end{equation}
where $S_{\mathrm{H}}$ is the Holst term (\ref{HolsttermBoundary}) without boundary term. 
Contrary to what happens to the Euler and Pontryagin terms, the Nieh-Yan term has a different asymptotic behavior,
it is finite, but not differentiable, for 
asymptotically flat spacetimes \cite{crv1}. Thus, even when it is by itself a boundary term, it has to be supplemented 
with an appropriate boundary term to make the variational principle well 
defined. This boundary term coincides precisely
with the boundary term in (\ref{HolsttermBoundary}) (up to a multiplicative constant), and the
resulting well defined Neih-Yan action is given by
\begin{equation} \label{well_defined_NY}
S_{\mathrm{NYB}}=S_{\ny}+\int _{\partial \mathcal{M}}  \Sigma^{IJ} \wedge \star\o_{IJ} \, .
\end{equation}


The equations of motion of the complete action (\ref{complete-action}), are the same as the ones
obtained from the Palatini action \cite{crv1},
\begin{eqnarray}
\label{EOMPalatiniFe} 
\varepsilon_{IJKL} e^{J} \wedge F^{KL} &=& 0\, , \\
\label{EOMPalatiniDe} \varepsilon_{IJKL} e^{K} \wedge D e^{L} &=& 0\, .
\end{eqnarray}
From (\ref{EOMPalatiniDe}) it follows that $T^L = 0$, and this is the 
condition of the compatibility of $\omega_{IJ}$ and $e^{I}$, that implies
\be
\o_{aIJ}=e^b_{[I}\partial_a e_{bJ]}+\Gamma^c_{ab}e_{c[I}e^b_{J]}\, ,
\ee
where $\Gamma^c_{ab}$ are the Christoffel symbols of the metric $g_{ab}=\eta_{IJ}e_a^Ie_b^J$.
Now, the equations (\ref{EOMPalatiniFe}) are equivalent to Einstein's equations $G_{ab}=0$.

\subsection{Boundary conditions}
\label{sec:3.2}

In this section we shall consider specific boundary conditions that are physically motivated. For the outer boundary we will specify 
asymptotically flat boundary conditions that capture the notion of isolated systems. For the inner boundary we will consider isolated horizons boundary conditions. In this way, we 
allow for the possibility of spacetimes that contain a black hole. This section has two parts. In the first one, we consider the outer boundary conditions and in the second part, 
the inner horizon boundary condition. 

\subsubsection{Asymptotically flat spacetimes}
\label{sec:3.3}

We are interested in spacetimes that at infinity look like a flat spacetime, 
in other words,  whose metric approaches a Minkowski metric at infinity 
(in some appropriately chosen coordinates). 
Here we will follow the standard definition of asymptotically flat spacetimes in the first order formalism 
(see e.g. \cite{aes}, \cite{cwe} and for a nice and pedagogical introduction in the metric formulation \cite{abr} and \cite{W}). 
Here we give a brief introduction into asymptotically flat spacetimes, following closely \cite{aes}.

In order to describe the behaviour of the metric at spatial infinity, we will focus on the region 
$\mathcal{R}$, that is the region outside the light cone of some point $p$.  We define a $4-$dimensional radial 
coordinate $\rho$ given by $\rho^{2} = \eta_{ab} x^{a} x^{b}$, where $x^{a}$ are the Cartesian coordinates 
of the Minkowski metric $\eta$ on $\mathbb{R}^{4}$ with origin at $p$. 
We will foliate the asymptotic region by timelike hyperboloids, $\mathcal{H}$, given by $\rho = \mathrm{const}$, 
that lie in $\mathcal{R}$. 
Spatial infinity $\mathcal{I}$ corresponds to a limiting hyperboloid when $\rho\to\infty$.
The standard angular coordinates on a hyperboloid are denoted by $\Phi^i =(\chi, \theta, \phi )$.

As shown in details in \cite{aes} and \cite{crv1} for asymptotically flat spacetime one can obtain the 
fall off conditions for tetrads and connection, and in order to have a well defined Lorentz angular momentum one needs to admit
an expansion of order 2, therefore we assume that in Cartesian coordinates we have the following behaviour
\begin{equation}\label{AFfalloff-tetrad}
e^{I} _{a} = \,^{o}e^{I} _{a} + \frac{\,^{1}e^{I} _{a}(\Phi )}{\rho} +  
\frac{\,^{2}e^{I} _{a}(\Phi )}{\rho^{2}} + o(\rho^{-2}),
\end{equation}
where $\,^{0}e^{I}$ is a fixed co-frame such that 
$g^{0}_{ab} = \eta_{IJ} \,^{o}e^{I} _{a} \,^{o}e^{I} _{b} $ is flat and $\partial_a\,^{o}e^{I}_b=0$. 

The sub-leading term $\,^{1}e^{I} _{a}$ is given by \cite{aes},
\begin{equation}\label{1e}
\,^{1}e^{I} _{a} = \sigma(\Phi) (2 \rho_{a} \rho^{I} - \,^{o}e^{I} _{a} ) 
\end{equation}
where
\begin{equation}\label{rhoa-rhoI}
\rho_{a} = \partial_{a} \rho \,\,\,\,\,\,\,\,   ,  \,\,\,\,\,\,\,\,   \rho^{I} = \,^{o} e^{aI} \rho_{a}.
\end{equation}
and $\sigma (-\chi ,\pi -\theta ,\phi +\pi )=\sigma (\chi ,\theta ,\phi )$. The last condition restrict the 
asymptotic behaviour of the metric, but is necessary in order
to reduce the asymptotic symmetries to a Poincar\'e group, as demonstrated in \cite{aes}.

The asymptotic expansion for connection can be obtained from the requirement that the connection be
compatible with tetrad on $\mathcal{I}$, to appropriate leading order. This leads to the asymptotic expansion 
of order 3 for the connection,
\begin{equation}\label{AFfalloff-connection}
\omega_{a} ^{IJ} = \,^{o} \omega_{a} ^{IJ} + \frac{ \,^{1} \omega_{a} ^{IJ}}{\rho} + 
\frac{ \,^{2} \omega_{a} ^{IJ}}{\rho^{2}} + \frac{ \,^{3} \omega_{a} ^{IJ}}{\rho^{3}} + o(\rho^{-3})\, .
\end{equation}

We require that $De^I$ vanishes, to an appropriate order, more precisely, we ask that the term of order 0 in $De^I$
vanishes
\begin{equation}
\d \,^{o}e^{I} + \,^{o} {\omega^I}_K\w \,^{o}e^{K}=0\, ,
\end{equation}
and since $ \d \,^{o}e^{I}=0$ it follows that $\,^{o} \omega^{IK}=0$. The term of order 1 should
also vanish leading to $\,^{1} \omega^{IK}=0$. We also ask that the term of order 2 in $De^I$
vanishes, and we obtain
\begin{equation}\label{compatibility}
\d \bigl( \frac{\,^{1}e^{I}}{\rho}\bigr) =-\frac{ \,^{2} {\omega^I}_K}{\rho^{2}}\w \,^{o}e^{K}\, ,
\end{equation}
and we shall demand compatibility between $e$ and $\o$ only based on these conditions.
As a result, we obtain 
$\,^{2} \omega_{a} ^{IJ} (\Phi) = 2 \rho^{2}\, \partial^{[J} (\rho^{-1} \,^{1}e^{I]} _{a} )$.
Note that although $\rho$ appears explicitly in the previous expression, it is independent of $\rho$.

Therefore, in the asymptotic region we have $De^I = O(\rho^{-3})$. This condition has
its repercussions on the behaviour of the Holst and Neih-Yan terms, as shown in \cite{crv1}.

\subsubsection{Internal boundary: Isolated horizons}
\label{sec:3.4}

A weakly isolated horizon is
a non-expanding null 3-dimensional hypersurface, with an additional condition 
that implies that surface gravity is constant on a horizon. Let us specify 
its definition and basic properties \cite{afk}.

Let $\Delta$ be a 3-dimensional null surface of $(\mathcal{M},g_{ab})$, equipped 
with future directed
null normal $l$. Let $q_{ab}\,\hat{=}\, g_{\underleftarrow{ab}}$ be the (degenerate)
induced metric on $\D$ (we denote by $\hat{=}$ an equality which holds only on $\D$
and the arrow under a covariant index denotes the pullback of a corresponding form to
$\D$). A tensor $q^{ab}$ that satisfies $q^{ab}q_{ac}q_{bd}\,\heq\, q_{cd}$, is called an
inverse of $q_{ab}$. The expansion of a null normal $l$ is defined by $\theta_{(l)}=
q^{ab}\nabla_a l_b$, where $\nabla_a$ is a covariant derivative compatible with the
metric $g_{ab}$. 

The null hypersurface $\D$ is called a {\it non-expanding horizon} if it satisfies the
following conditions: (i) $\D$ is topologically $S^2\times\re$, (ii) $\theta_{(l)}=0$
for any null normal $l$ and (iii) all equations of motion hold at $\D$  and 
$-T_{ab}l^b$ is future directed and causal for any $l$, where 
$T_{ab}$ is matter stress-energy tensor at $\D$. The second condition implies that
the area of the horizon is constant `in time', so that the horizon is isolated. 

We need one additional condition in order to satisfy the zeroth law of black hole
dynamics. In order to introduce it let us first specify some details of the geometry
of the isolated horizon. It is convenient to use null-tetrads $(l,n,m,\bar{m})$, where
a real, future directed null vector field $n$ is transverse to $\D$ and a complex vector
field $m$ is tangential to $\D$, such that $l\cdot n=-1$, $m\cdot\bm =1$ and all the
other scalar products vanish.

Since $l$ is a null normal to $\D$ it is geodesic and its twist vanishes. We define
surface gravity $\k_{(l)}$ as the acceleration of $l^a$
\be
l^a\nabla_a l^b\, \heq\, \k_{(l)}l^b\, .
\ee

The Raychaudhuri and Einstein's equations together with the condition on the 
stress-energy tensor imply that every $l$ is also shear free and since its
expansion and twist vanish there exists a one-form $\o_a$ such that \cite{chandra}
\be
\nabla_{\ula{a}}l^b\, \heq\, \o_a l^b\, .\label{5.1}
\ee


Since $l$ can be rescaled by an arbitrary positive function, in general $\k_{(l)}$ is not 
constant on $\D$. If we want to establish the zeroth law of black hole dynamics
$\d\k_{(l)}\,\heq\, 0$ we need one additional condition, the `time' invariance of $\o$,
\begin{equation}
 \La_l\o\,\heq\, 0\, .
\end{equation}
Now, if we restrict to constant rescaling of $l$, $l\to l'=cl$ that leaves
$\o$ invariant, then the zeroth law of black hole dynamics follows,
for every null normal $l$ related to each other by constant rescaling.

All null normals related to each other by a constant rescaling form an equivalence class 
$[l]$. Now, we can define a {\it weakly isolated horizon} (WIH) $(\D ,[l])$ as
a non-expanding horizon equipped with an equivalence class $[l]$, such that
$\La_l\o\,\heq\, 0$, for all $l\in [l]$.

In order to analyze the contribution to the variation of the action over the internal 
boundary, which is a WIH $\D$, we equip $\D$ with a fixed class of null normals $[l]$
and fix an internal null tetrads $(l^I,n^I,m^I,\bm^I)$ on $\D$, such that their
derivative with respect to flat derivative operator $\partial_a$ vanishes.The
permissible histories at $\D$ should satisfy two conditions: (i) the vector field
$l^a:=e^a_Il^I$ should belong to the fixed equivalence class $[l]$ (this is a
condition on tetrads) and (ii) the tedrads and connection should be such that $(\D ,[l])$
constitute a WIH.

The expression for tetrads on $\D$ is given by \cite{cg}
\be
e_a^I\,\heq \, -l^I n_a + \bm^I m_a +m^I\bm_a\, ,
\ee
so that
\be
\Sigma^{IJ}\,\heq\, 2l^{[I}n^{J]}\,\eps +2i\, n\w (m\, l^{[I}\bm^{J]}-\bm\, l^{[I}m^{J]})\, ,\label{hor1}
\ee
where we introduced the area two-form on the cross-sections
of $\D$, $\eps := im\w\bm$, that is also preserved in `time', $\La_l\eps\,\heq\, 0$ \cite{afk}.
The expression for the connection on $\D$ is given by \cite{cg}
\be
\o_{IJ}\,\heq\, -2\o\, l_{[I}n_{J]} + 2U\, l_{[I}\bm_{J]} + 2\bar{U}\, l_{[I}m_{J]}
+ 2V\, m_{[I}\bm_{J]}\, ,\label{hor2}
\ee
where we have introduced two new one-forms, a complex one $U$ and purely imaginary one $V$.
In \cite{cg} the expression for these one forms is given in terms of Newman-Penrose (NP)
spin coefficients and null tetrads. Here, we will only state the results that we will
need in the following sections (details can be seen in \cite{afk}, \cite{crv1} and \cite{cg} ).
It turns out that $l\cdot \d\o\,\heq\, l\cdot \d V\,\heq\, 0$ and we can use the residual Lorentz 
transformations on the horizon in order to obtain $l\cdot V\,\heq\, 0$.

We shall also need the expression for the pull-back of the curvature two-form on the cross-section of a weakly
isolated non-rotating horizon (the details are given in \cite{liko_booth}), it follows that
\be
2 m_{[I}\bar{m}_{J]} F^{IJ}\vert_{S_\D}=-i\,\mathcal{R}\,\eps\,,\label{pullback_crossection}
\ee
where $\mathcal{R}$ is the scalar curvature of the cross-section of $\Delta$.

\section{Conserved charges}
\label{sec:4}
In this section we shall consider some of the information that comes from the covariant Hamiltonian formulation. 
In particular, we shall see how one can define conserved quantities. As we have discussed in Secs.~\ref{sec:2.2} 
and \ref{sec:2.3} there are two classes of quantities, namely those that are generators of Hamiltonian symmetries 
and the so called Noether charges. We shall then analyze the relation between Hamiltonian and Noether charges
for the most general first order gravitational action, focusing on the role that the boundary terms play.
As one might anticipate, the fact that the boundary terms do not modify the symplectic structure implies that 
the Hamiltonian charges are insensitive to the existence of extra 
boundary terms. However, as we shall see in detail, the Noetherian quantities {\it do} depend on the boundary terms.
Specifically, we are interested in the relation of the Noether charge with the energy at the asymptotic region and the
energy of the horizon. 

\subsection{Hamiltonian charges}
\label{sec:4.1}

From equations (\ref{sympl_poten_boundary_term}) and (\ref{PalatiniplusBoundary}), the symplectic potential 
for the well posed Palatini action $S_{\mathrm{PB}}$ is given by
\begin{equation}\label{SympPotentialPalatini}
\Theta_{\mathrm{PB}} (\delta ) = 
 \frac{1}{2 \kappa} \int_{\partial \mathcal{M}} \delta \Sigma^{IJ} \wedge \omega_{IJ}\, .
\end{equation}
Therefore from (\ref{defJ}) and (\ref{SympPotentialPalatini}) the corresponding symplectic current is,
\begin{equation}\label{JPalatini}
J_{\mathrm{P}}(\delta_{1}, \delta_{2})  = -\frac{1}{2 \kappa} \left( \delta_{1} \Sigma^{IJ} \wedge \delta_{2} \omega_{IJ} 
- \delta_{2} \Sigma^{IJ} \wedge \delta_{1} \omega_{IJ} \right)\, .
\end{equation}
Note that the symplectic current  
is insensitive to the boundary term, as we discussed in Sec.~\ref{sec:2}.
From the equation (\ref{intJzero}) one can obtain a conserved
pre-symplectic structure, as an integral of $J_{\mathrm{P}}$ over a spatial surface, if the integral
of the symplectic current over the asymptotic region vanishes and if the integral over an isolated horizon
behaves appropriately. 
As shown in \cite{aes}, for asymptotically flat spacetimes, $\int_{\mathcal{I}}J_{\mathrm{P}}=0$, and on a  WIH
we have $J_{\mathrm{P}}\,\heq\, \d j$, \cite{afk}.  As a result,
the conserved pre-symplectic structure for the Palatini action,
for asymptotically flat spacetimes with weakly isolated horizon, takes the form \cite{afk}
\begin{equation}
\bar{\O}_P(\de_1,\de_2)=-\frac{1}{2\k}\int_{M} \left( \de_1\S^{IJ}\w\de_2\o_{IJ} - \de_2\S^{IJ}\w\de_1\o_{IJ}\right)
-\frac{1}{\k}\int_{S_\D}\de_1\psi\,\de_2 (\eps )-\de_2\psi\,\de_1 (\eps )\, ,\label{pal_ss}
\end{equation}
where $S_{\D}$ is a 2-sphere at the intersection of a Cauchy surface $M$ with a horizon and 
$\psi$ is a potential defined as
\be
\La_l\psi =\k_{(l)}\, ,\ \ \ \psi =0\ \ {\rm on}\ \ S_{1\D}\, ,\label{psi}
\ee
with $S_{1\D}=M_1\cap\D$. 
We see that the existence of an isolated horizon modifies the symplectic structure of the theory. 

Let us now see what is the contribution of the well posed Holst term, $S_{\mathrm{HB}}$. 
In this case the symplectic potential is given by \cite{cwe}
\be
\Theta_{\mathrm{HB}} (\de )=\f {1}{2\k\g}\int_{\partial \mathcal{M}} \de\S^{IJ}\w\star\o_{IJ}=
\f {1}{\k\g}\int_{\partial \mathcal{M}} \de e^I\w\d e_I\, ,
\ee
where in the second line we used the equation of motion $De^I=0$. The symplectic current in this case is a total derivative and is given by
\begin{equation}\label{symplectic_current_holst}
J_{\mathrm{H}}(\delta_{1}, \delta_{2})  = \frac{1}{\k\g}\d\, (\de_1 e^I\w\de_2 e_I)\, .
\end{equation}
As we have seen in the Sec. \ref{sec:2}, when the symplectic current is a total derivative, the covariant
Hamiltonian formalism indicates that the corresponding (pre)-symplectic structure vanishes.

As we also remarked in Sec.~\ref{sec:2}, one could postulate a conserved two form $\tilde\O$
if $\int_{\mathcal I}J_{\mathrm{H}}=0$ and $\int_{\D}J_{\mathrm{H}}=0$, in which case
this term defines a conserved symplectic structure. Let us, for completeness, consider this
possibility. In \cite{cwe} it has been shown that the integral at ${\mathcal I}$
vanishes, so here we shall focus on the integral over $\Delta$
\begin{equation}\label{Holst_int_horizon}
 \int_{\D}J_{\mathrm{H}}=\frac{1}{\k\g}\int_{\partial\D}\de_1 e^I\w\de_2 e_I=\frac{1}{\k\g}\int_{\partial\D}
 \de_1 m\w \de_2\bm + \de_1\bm\w\de_2 m\, .
\end{equation}
We can perform an appropriate Lorentz transformation at the horizon in order to get a foliation of $\D$ spanned
by $m$ and $\bm$, that is Lie dragged along $l$ \cite{afk}, that implies $\La_l m^a\,\heq\, 0$. At the other hand,
$\partial\D =S_{\Delta 1}\cup S_{\Delta 2}$, so it is sufficient to show that the integrand in (\ref{Holst_int_horizon})
is Lie dragged along $l$. The variations in (\ref{Holst_int_horizon}) are tangential to $S_{\D}$, hence we have
$\La_l\de_1 m = \de_1\La_l m=0$, so that the integrals over $S_{\Delta 1}$ and $S_{\Delta 2}$ are equal and
$\int_{\D}J_{\mathrm{HB}}=0$. So we can define a conserved pre-symplectic structure corresponding to the Holst term
\begin{equation}
 \tilde{\O}_H(\de_1,\de_2)=\frac{1}{\k\g}\int_{\partial M}\de_1 e^I\w\de_2 e_I\, ,
\end{equation}
where the integration is performed over $\partial M=S_{\infty}\cup S_{\D}$. As shown in \cite{cwe}, the integral over
$S_{\infty}$ vanishes, due to asymptotic conditions, and the only contribution comes from $S_{\D}$. Finally, we see that the quantity
\begin{equation}\label{symplectic_structure_holst}
 \tilde{\O}_H(\de_1,\de_2)=\frac{1}{\k\g}\int_{S_{\D}}\de_1 e^I\w\de_2 e_I\, .
\end{equation}
defines a conserved two-form. Note that this is precisely 
the symplectic structure for the Holst term defined in \cite{merced-holst}, though there the authors did not explicitly show
that it is independent of $M$ (this result depends on the details of the boundary conditions).

As we have seen in (\ref{defJ}) the boundary terms in the action (that is, the topological terms) do not contribute to the symplectic current $J$, 
so that the only contributions in our case come 
from the Palatini action and possibly, as we have just seen, 
from the Holst term\footnote{Note that there have been some
statements in the literature claiming that the topological terms do contribute to the symplectic structure when there are boundaries present \cite{liko_2, m:m, perez_2}.}. 
In order to illustrate how some possible inconsistencies arise when one postulates the existence of a symplectic structure for the topological terms, let us see, with some detail, what happens in the case of the Pontryagin term (as suggested, for instance, in \cite{m:m}. Similar results follow for the other topological terms.) Recall that this term can be written as a total derivative, which means that we can either view it as a bulk term or as a boundary term. Considering the derivation of the symplectic structure in either case should render equivalent descriptions. Let us consider the
 variation of $S_{\Po}$, calculated from the LHS (bulk expression) in (\ref{Pontryagin}), is
\be
\de S_{\Po}=-2 \int_{\mathcal{M}} DF^{IJ}\w\de\o_{IJ}+
2\int_{\partial \mathcal{M}}F^{IJ}\w\de\o_{IJ}\, ,\label{var_Po1a}
\ee
so it does not contribute to the equations of motion in the bulk, due to the Bianchi
identity $DF^{IJ}=0$.
In this case, the corresponding symplectic current is 
\begin{equation}
J_{\Po}^{\rm bulk}(\de_1,\de_2)=2(\de_1 F^{IJ}\w\de_2\o_{IJ} -  \de_2 F^{IJ}\w\de_1\o_{IJ})\, .
\end{equation}
On the other hand, if we calculate the variation of the Pontryagin term directly from the 
RHS (boundary expression) of (\ref{Pontryagin}), we obtain
\begin{equation}\label{var_Po2}
\de S_{\Po}=2 \int_{\partial \mathcal{M}}\de L_{CS}\, .
\end{equation}
The two expressions for $\de S_{\Po}$ are, of course, identical since
$F^{IJ}\w\de\o_{IJ}=\de L_{CS}+\d (\o^{IJ}\w\de\o_{IJ})$. 
The corresponding symplectic current in this case is
\begin{equation}
J_{\Po}^{\rm bound}(\de_1,\de_2)=4\,\de_{[2}\de_{1]} L_{CS}=0\, ,
\end{equation}
as we have obtained in the Sec.~\ref{sec:2}. So, at first sight it would seem that there is an ambiguity in the definition
of the symplectic current that could lead to different symplectic structures. 
Since the relation between them is given by
\begin{equation}
 J_{\Po}^{\rm bulk}(\de_1,\de_2)=J_{\Po}^{\rm bound}(\de_1,\de_2)+4\,\d (\de_2\o^{IJ}\w\de_1\o_{IJ})\, ,
\end{equation}
it follows that $J_{\Po}^{\rm bulk}(\de_1,\de_2)$ is a total derivative, that does not contribute in
(\ref{intJzero}), and from the systematic derivation of the symplectic structure described in Sec.~\ref{sec:2}, we have to conclude that it does not contribute to the symplectic structure. 
This is consistent with the fact that $J_{\Po}^{\rm bound}$
and $J_{\Po}^{\rm bulk}$ correspond to the same action. As we have remarked in  Sec.~\ref{sec:2}, a total
derivative term in $J$, under some circumstances, could be seen as generating a non-trivial symplectic
structure $\tilde\O$ on the boundary of $M$. But the important thing to note here is that one would run into an inconsistency if one choose to introduce that non-trivial $\tilde\O$. Thus, consistency of the formalism requires that $\tilde\O=0$.

Let us now construct the conserved charges for this theory, and from the previous reasons we shall only consider the Palatini and Holst terms in this part.
We shall consider the Hamiltonian $H_\xi$ that is a conserved quantity corresponding
to  asymptotic symmetries and symmetries on the horizon of a spacetime. Our asymptotic conditions are
chosen in such a way that the asymptotic symmetry group be the Poincar\'e group. The corresponding
conserved quantities, for the well posed Palatini action, energy-momentum and relativistic angular momentum, are constructed in \cite{aes}.
The contribution to the energy from a weakly isolated horizon has been analyzed in \cite{afk}, where the 
first law of mechanics of non-rotating black holes was deduced. Rotating isolated horizons have been
the topic of \cite{abl}, where the contribution from the angular momentum of a horizon has
been included. In this paper we restrict our attention to energy and give a review of the principal results presented in \cite{afk}.

Let us consider a case when $\xi$ is the infinitesimal generator of asymptotic time 
translations of the spacetime. It induces time evolution
on the covariant phase space, generated by a vector field $\de_\xi :=(\La_\xi e,\La_\xi\o )$.
At infinity $\xi$ should approach a
time-translation Killing vector field of the asymptotically flat spacetime. On the 
other hand, if we have a non-rotating horizon $\D$, then $\xi$, at the horizon, should belong to the 
equivalence class $[l]$. In order that $\de_\xi$ represents a phase space symmetry the condition
$\La_{\de_\xi}\bar{\O} =0$ should be satisfied. As we have seen in Sec.~\ref{sec:2.2}, $\de_\xi$ is a Hamiltonian vector field iff the one-form
\be
X_\xi (\de )=\bar{\O} (\de ,\de_\xi )\, ,
\ee
is closed, and the Hamiltonian $H_\xi$ is defined as
\be
X_\xi (\de )=\de H_\xi\, .
\ee
In the presence of the isolated horizon, the symplectic structure (\ref{pal_ss}) has two contributions, 
one from the Cauchy surface $M$ and the 
other one from the two-sphere $S_\D$. This
second term does not appear in $\bar{\O}_P (\de ,\de_\xi )$, it is equal to
\be
-\frac{c}{\k}\int_{S_\D} \de_l\eps\,\de\psi -\de\eps\,\de_l\psi\, ,\label{int}
\ee
since $\xi=cl$ at $\D$. We will show that this integral vanishes. 
When acting on fields $\de_l =\La_l$, so that $\de_l\eps =\La_l\eps\,\heq\, 0$. On the 
other hand, as pointed out in \cite{abl}, $\psi$ is a potential, a given
function of the basic variables. If we define $\de_l\psi =\La_l\psi$, as in the case
of our basic fields, then the boundary condition in (\ref{psi}) cannot be fulfilled.
So we need to define $\de_l\psi$ more carefully.
Let $\psi '$ denote a potential that corresponds to
a null vector $l'=cl$ in the sense that $\La_{l'}\psi '=\k_{(l')}$. Since 
$\k_{(l')}=c\k_{(l)}$, it follows that $\La_l\psi '=\k_{(l)}$. We define 
$\de_l\psi =\psi '-\psi$, then
\be
\La_l(\de_l\psi )=0\, .
\ee
We ask $\psi$ to be fixed at $S_{1\D}$, as a result $\de_l\psi\,\heq\, 0$. Then, it follows
that the integral (\ref{int}) vanishes and the only contribution to 
$\bar{\O}_P (\de ,\de_\xi )$ comes from the integral over the Cauchy surface $M$ in (\ref{pal_ss}).

On the other hand, the symplectic structure for the Holst term $\tilde{\O}_H$ (\ref{symplectic_structure_holst}) 
is restricted to $S_{\Delta}$, but it turns out that
\be
\tilde{\O}_H (\de ,\de_\xi )=\frac{c}{\k\g}\int_{S_{\D}}\de m\w \La_l \bm + \de\bm\w\La_l m=0\, .
\ee
As a result $\de H_\xi := \bar{\O} (\de ,\de_\xi )=\bar{\O}_P(\de ,\de_\xi )$ only has a contribution from the Palatini action 
\be
\de H_\xi =-\frac{1}{2\k}\int_{\partial M}(\xi\cdot\o^{IJ} )\de\S_{IJ} -
(\xi\cdot\S_{IJ} )\w\de\o^{IJ} \, ,\label{firstlaw}
\ee
where the integration is over the boundaries of the Cauchy surface $M$,
the two-spheres $S_\infty$ and $S_\D$, since the integrand in 
$\bar{\O}_P (\de ,\de_\xi )$ is a total derivative, 
as shown in \cite{afk}.

The asymptotic
symmetry group is the quotient of the group of spacetime diffeomorphisms which preserve the 
boundary conditions by its subgroup consisting of asymptotically identity diffeomorphisms. In our case 
this is the Poincar\'e group and its action generates canonical transformations on the covariant
phase space whose generating function is $H_\xi^\infty$. The situation is similar at the horizon $\D$
and infinitesimal diffeomorphisms need not be in the kernel of the symplectic structure unless they vanish
on $\D$ and the horizon symmetry
group is the quotient of the Lie group of all infinitesimal spacetime diffeomorphisms
which preserve the horizon structure by its subgroup consisting of elements which are identity
on the horizon \cite{abl}.

The surface term at infinity in the expression (\ref{firstlaw}) defines the gravitational energy at
the asymptotic region, whose variation
is given by
\begin{equation}\label{energy_infinity}
\de E^\xi_\infty := -\frac{1}{2\k}\int_{S_\infty}(\xi\cdot\o^{IJ} )\de\S_{IJ} -
(\xi\cdot\S_{IJ} )\w\de\o^{IJ} = \frac{1}{2\k}\int_{S_\infty}(\xi\cdot\S_{IJ} )\w\de\o^{IJ}\, ,
\end{equation}
since, due to the asymptotic behaviour of the tetrad and connection, the first term in the above
expression vanishes. As shown in \cite{aes}, after inserting the asymptotic form of the tetrad 
(\ref{AFfalloff-tetrad}) and connection (\ref{AFfalloff-connection}), this integral represents
the variation of the ADM energy, $\de E^\xi_{\mathrm{ADM}}$, associated with the asymptotic 
time-translation defined by $\xi$
\begin{equation}
 E^\xi_\infty = E^\xi_{\mathrm{ADM}}=\frac{2}{\k}\int_{S_\infty}\sigma\,\d^2 S_o\, ,
\end{equation}
where $\d^2 S_o$ is the area element of the unit 2-sphere. 

On the other hand, the surface term at the horizon in the expression (\ref{firstlaw}) represents the  
horizon energy defined by the time translation $\xi$, whose variation is given by
\begin{equation}\label{energy_horizon}
\de E^\xi_\D := \frac{1}{2\k}\int_{S_\D}(\xi\cdot\o^{IJ} )\de\S_{IJ} -
(\xi\cdot\S_{IJ} )\w\de\o^{IJ} = \frac{1}{2\k}\int_{S_\D}(\xi\cdot\o^{IJ} )\de\S_{IJ}\, ,
\end{equation}
since the second term in the above expression vanishes at the horizon. The remaining term is
of the form
\begin{equation}
\de E^\xi_\D =  \frac{1}{\k}\,\k_{(\xi )}\de a_\D\, ,
\end{equation}
where $a_\D$ is the area of the horizon.

Now we see that the expression 
 (\ref{firstlaw}) encodes the first law of mechanics for non-rotating black holes, since
it follows that
\be\label{first law}
\de H_\xi =\de E^\xi_{\mathrm{ADM}}-\frac{1}{\k}\,\k_{(\xi )}\de a_\D\, .
\ee
We see that the necessary condition for the existence
of $H_\xi$ is that surface gravity, $\k_{(\xi )}$, be a function only of a horizon area $a_\D$.
In that case
\be
H_\xi =E^\xi_{\mathrm{ADM}}-E^\xi_\D\, .
\ee

In the following section we want to calculate the Noether charge that corresponds to time translation
for every term of the action (\ref{complete-action}). We have just seen that $\de H_\xi$ is
an integral over a Cauchy surface of the symplectic current $J(\de ,\de_\xi )$.
In section~\ref{sec:2.3} we displayed the relation between the symplectic and Noether currents,
given in (\ref{nc}), and using the definition of Noether charge $Q_\xi$ (\ref{nch}), we obtain the
following relation
\be
\de H_\xi = \int_M J(\de ,\de_\xi ) = \int_{\partial M}\de Q_\xi -\xi\cdot\theta (\de )\, .\label{int2}
\ee
There are two contributions to the above expression, one at $S_\infty$ and the other
one at $S_\D$. As before, $\de E_{\infty}^\xi$, is the integral at the RHS of (\ref{int2}) calculated over
$S_\infty$, and  $\de E_{\D}^\xi$ the same integral calculated over $S_\D$.
Note that the necessary and sufficient condition for the existence of $H_\xi$ is the existence
of the form $B$ such that
\be
\int_{\partial M}\xi\cdot\theta (\de )=\de \int_{\partial M}\xi\cdot B\, .\label{condB}
\ee
Let us now consider how the different terms appearing in the action contribute to the Noether charges.

\subsection{Noether charges}
\label{sec:4.2}

In this part we consider the Noether charges that appear as conserved quantities associated to
diffeomorphisms generated by vector fields $\xi$. There are two parts. In the first one we consider
in detail the consistent Palatini action with a boundary term, and compare it to the case without
a boundary term. In the second part we consider the Holst and topological terms.

\subsubsection{Palatini action}

Let us start by considering the case of Palatini action with boundary term. We have seen in the
section~\ref{sec:4.1} that the symplectic potential current in this case is given by
\be
\theta_{\mathrm{PB}}(\de )=\frac{1}{2\k }\,\de\S^{IJ}\w\o_{IJ}\, .
\ee
In order to calculate  the Noether current 3-form (\ref{noether}), $J_N(\de_\xi )=\theta (\de_\xi )-\xi\cdot {\mathbf L}$, 
we need the following two expressions
\be
\theta_{\mathrm{PB}} (\de_\xi )=\frac{1}{2\k}\,\La_\xi\S^{IJ}\w\o_{IJ}=\frac{1}{2\k}\, 
[\d (\xi\cdot\S^{IJ})+\xi\cdot\d\S^{IJ}]\w\o_{IJ}\, ,
\ee
and
\be
\xi\cdot {\mathbf L}_{\mathrm{PB}}=-\frac{1}{2\k}\, [\xi\cdot (\S^{IJ}\w F_{IJ})- \xi\cdot\d (\S^{IJ}\w\o_{IJ})]\, .
\ee
From these expressions we obtain the following result for the Noether current 3-form
\be
J_{\mathrm{NPB}}(\de_\xi ) = \frac{1}{2\k}\, \{(\xi\cdot\S^{IJ})\w F_{IJ} + (\xi\cdot\o_{IJ}){\rm D}\S^{IJ} + 
\d [(\xi\cdot\S^{IJ})\w\o_{IJ}]\}\, .
\ee
We see that on-shell ($e^I\w F_{IJ}=0$ and ${\rm D}\S^{IJ}=0$), we have $J_N(\de_\xi )=\d Q_\xi$, where
the corresponding Noether charge is given by
\be
Q_{\xi\mathrm{PB}} =\frac{1}{2\k}\, (\xi\cdot\S^{IJ})\w\o_{IJ}\, .
\ee
We shall show that the contribution of the second term in (\ref{int2}) over $S_\infty$ vanishes. Namely,
\be
\int_{S_\infty}\xi\cdot\theta_{\mathrm{PB}} (\de )=\frac{1}{2\k}\,\int_{S_\infty}\xi\cdot (\de\S^{IJ}\w\o_{IJ})=0\, ,
\ee
since $\de\S = O({\rho^{-1}})$, $\o = O({\rho^{-2}})$ and the volume element goes as $\rho^2$.
It follows that $B=0$ on $S_\infty$ and the Hamiltonian at infinity exists.
The remaining term at infinity in (\ref{int2}) is
\be
\de\int_{S_\infty}Q_{\xi\mathrm{PB}} =\frac{1}{2\k}\,\de\int_{S_\infty}(\xi\cdot\S^{IJ})\w\o_{IJ}\, ,
\label{adm}
\ee
and since $\int_{S_\infty}\de (\xi\cdot\S^{IJ})\w\o_{IJ}=0$, due to asymptotic behaviour of the fields,
the above expression is equal to $\de E_\infty^\xi$ given in (\ref{energy_infinity}), so in this case
\be
E^\xi_{\rm ADM}=\int_{S_\infty}Q_{\xi\mathrm{PB}}\, ,
\ee
up to an additive constant that we choose to be zero. Note that a similar result is obtained in the second 
order formalism for the Einstein-Hilbert action with the Gibbons-Hawking term, as
shown in \cite{pons}.

On the other hand, at the horizon the situation is different. In fact,
\be
\int_{S_\D}Q_{\xi\mathrm{PB}} =\frac{1}{2\k}\,\int_{S_\D}(\xi\cdot\S^{IJ})\w\o_{IJ}=0\, ,
\ee
because $\xi^a= cl^a$ on the horizon, and due to the expressions for $\Sigma$ (\ref{hor1}) and $\o$ (\ref{hor2}) at the horizon.
Then, $\de E_{\D}$ is determined by the remaining term
\be
\int_{S_\D}\xi\cdot\theta_{\mathrm{PB}} (\de )
=\frac{1}{2\k}\,\int_{S_\D} [(\xi\cdot\de\S^{IJ})\w\o_{IJ}+ \de\S^{IJ} (\xi\cdot\o_{IJ})]\, .
\ee
The first term vanishes since on $\D$ we have 
$(\xi\cdot\de\S^{IJ})\w\o_{IJ}\,\heq\, 2c(l\cdot\eps )\w\o\,\heq\, 0$, because $l\cdot\eps\,\heq\, 0$.
We are then left with the expression given in (\ref{energy_horizon}),
and the necessary condition for the existence of $E_{\D}$ is that the surface gravity
$\k_{(\xi )}$ depends only on the area of the horizon \cite{afk}. It follows also that
there exists a form $B$ such that (\ref{condB}) is satisfied. 

We see that in this case
\be
\de E_{\infty}=\de\int_{S_\infty}Q_{\xi\mathrm{PB}}\, ,\ \ \ \ \ 
\de E_{\D}=\int_{S_\D}\xi\cdot\theta_{\mathrm{PB}} (\de )\, .
\ee
In globally stationary spacetimes, $\La_\xi e =\La_\xi\o =0$, so that 
$\de H_\xi =\bar\O (\de ,\de_\xi )=0$, and from the first law (\ref{first law}) it follows $\de E_{\infty}=\de E_{\D}$.
For Palatini action with boundary term this implies that
\be
\de\int_{S_\infty}Q_{\xi\mathrm{PB}}= \int_{S_\D}\xi\cdot\theta_{\mathrm{PB}} (\de )\, .
\ee
This result depends on the particular form of the action, and it is sensitive to the presence of boundary terms.

Let us briefly comment the case of Palatini action {\it without} boundary term. We know that
this action is not well defined, its symplectic potential $\Theta_{\mathrm{P}}(\de )$ diverges,
but we can {\it formally}
calculate its Noether charge and compare it to the previous example. 
As we showed in the sections \ref{sec:2.1} and \ref{sec:2.3}, the addition of the total derivative to the action changes
its Noether charge (\ref{noether_charge_change}), but leaves the symplectic structure 
(\ref{defJ}) unaltered. In the previous example
the ADM energy was determined completely by the integral of the Noether charge over the two-sphere
at infinity. Now, the situation is different and both terms in (\ref{int2}) contribute to $\de E_\infty$.
We first note that
\be
\theta_{\mathrm{P}}(\de )=-\frac{1}{2\k}\, \S^{IJ}\w\de\o_{IJ}\, , \ \ \ \ 
Q_{\xi\mathrm{P}} = -\frac{1}{2\k}\, \S^{IJ}(\xi\cdot\o_{IJ})\, ,
\ee
where $\theta_{\mathrm{P}}$ and $Q_{\xi\mathrm{P}}$ denote the corresponding quantities for Palatini action (without boundary term). It turns out that
\be
\int_{S_\infty}\xi\cdot \theta_{\mathrm{P}}(\de )=-\frac{1}{2\k}\,\de\int_{S_\infty}\xi\cdot 
(\S^{IJ}\w\o_{IJ})\, ,
\ee
since $\int_{S_\infty}\xi\cdot(\de\Sigma^{IJ}\w\o_{IJ})=0$ due to our asymptotic conditions.
On the other hand,
\be
\int_{S_\infty}\de Q_{\xi\mathrm{P}} = -\frac{1}{2\k}\,\de\int_{S_\infty}(\xi\cdot\o_{IJ})\S^{IJ}
=   -\frac{1}{2\k}\int_{S_\infty} \de (\xi\cdot\o_{IJ})\S^{IJ}     \, ,
\ee
and the combination of the above expressions, as in (\ref{int2}) gives the previous expression
for $\de E^\xi_\infty$ (\ref{energy_infinity}). Thus, we see that the Hamiltonian generator at infinity is 
not given by the integral of the Noether charge, as in the case of the Palatini action with boundary term.

At the horizon both terms contribute, again. The results are
\be
\int_{S_\D}\xi\cdot \theta_{\mathrm{P}}(\de )=-\frac{1}{\k}\,\de\k_{(\xi )}a_\D\, ,
\ee
where we used the fact that $l\cdot\eps =0$ and $\xi\cdot\de\o =\de (\xi\cdot\o )=\de\k_{(\xi )}$.
We see again, that in order to satisfy the condition (\ref{condB}), $\k_{(\xi )}$
should be a function of $a_\D$ only. We also obtain
\be\label{variation_noether_charge_horizon_palatini}
\de\int_{S_\D} Q_{\xi\mathrm{P}}=\frac{1}{\k}\,\de (\k_{(\xi )}a_\D )\, ,
\ee
and the combination of the above expressions, as in (\ref{int2}) gives the previous result
for $\de E^\xi_\D$ (\ref{energy_horizon}).

Finally, let us compare these results for the Noether charge with the results of \cite{iw},
and to that end we shall recall 
one of the principal results in \cite{iw}, referring to the variations of a stationary
black hole solution, that states that in diffeomorphism invariant theories, in the second order formalism, 
the Noether charge
relative to a bifurcate Killing horizon $\Sigma_0$ is proportional to the entropy of a black hole $S$.
The result  is the following
\begin{equation}\label{entropy Wald}
\de\int_{\Sigma_0} Q_{\xi_0} =\frac{\kappa_{(\xi_0)}}{2\pi}\de S\, , 
\end{equation}
where $\xi_0$ is the Killing field that vanishes on $\Sigma_0$ and at infinity tends to a stationary time-like
Killing vector field with unit norm and $\kappa_{(\xi_0)}$ is the corresponding surface gravity of a stationary
black hole. In the proof of this result it is assumed that $\de\kappa_{(\xi_0)} =0$. Furthermore, it has
been shown that in the case of stationary variations the integral is independent of the choice of horizon cross-section.
Our analysis, based on the IH formalism \cite{aes} and \cite{afk},
is different in various aspects: (1) we consider the first order formalism; (2) in our case the 
existence of the internal boundary is consistently treated, as, for example, in the expression
for $\de H$ that involves integration over the whole boundary, not only 
over the asymptotic region, as in \cite{iw}; (3) our results are valid also for nonstationary
configurations, and; (4) in our approach the integration is performed over an arbitrary 2-sphere cross section of
a weakly isolated horizon, and not restricted to a preferred bifurcation surface.

Taking this into account let us now see whether, in our approach, 
the Noether charge can be related to the black hole energy (or entropy).
We already know that in general this is not the case, since neither the Holst term nor the topological 
terms contribute to the energy of the black hole, though they modify the Noether charge.

We can formally compare the expression (\ref{entropy Wald}) with our result 
(\ref{variation_noether_charge_horizon_palatini}), taking into account all differences between the two
approaches. We see that, if we impose that $\de\k_{(\xi )} =0$, then the result in (\ref{variation_noether_charge_horizon_palatini})
would look like  (\ref{entropy Wald}). But this restriction is not consistent with the result of \cite{afk} that shows that,
as we saw in the previous subsection, the surface gravity is a function of the area
of the horizon, and that this is a necessary condition to have a well defined Hamiltonian. As we have seen in this subsection, in neither of the cases, namely Palatini action with 
or without surface term, is the variation
of a corresponding Noether charge relative to an isolated horizon proportional to $(\kappa_{(\xi )}\de a_\D)$.\footnote{nor to $\delta a_\D$, for that matter.} Note that this fact 
poses a challenge to the generality of the result relating Noether charge and energy (or entropy) derived in \cite{iw}. 

\subsubsection{Holst and topological terms}
\label{sec:4.3}

To end this section, let us calculate the Noether charges for the Holst term and the topological terms.
We shall see that in all of these cases the integrals of the corresponding Noether charge 2-form
over $S_\infty$ vanish. For $S_\D$, there is one case where the charge is non vanishing.
Let us first consider the Holst term with its boundary term $S_{\mathrm{HB}}$,
given by (\ref{HolsttermBoundary}). We know that this term does not contribute to the energy.
As we have seen in section~\ref{sec:4.2}, the symplectic potential current of $S_{\mathrm{HB}}$ is given by
\be
\theta_{\mathrm{HB}} (\de )=\f {1}{2\k\g}\de\S^{IJ}\w\star\o_{IJ}=\f {1}{\k\g}\de e^I\w\d e_I\, ,
\ee
where in the second line we used the equation of motion ${\rm D}e^I=0$. The corresponding Noether charge 2-form
is given by
\be
Q_{\xi\mathrm{HB}} = \f {1}{\k\g}(\xi\cdot\S^{IJ})\w\star\o_{IJ}=\f {1}{\k\g}(\xi\cdot e^I)\, \d e_I\, .
\ee
Now, one can show that
\begin{equation}
\int_{S_\infty} Q_{\xi\mathrm{HB}} = 0\, .
\end{equation}
Namely
\begin{equation}
\int_{S_\infty}(\xi\cdot e^I)\, \d e_I =
\int_{S_\infty}(\xi\cdot \,^{o}e^I)\, \d \bigl( \frac{\,^{1}e_I}{\rho}\bigr) =
\int_{S_\infty}\d \bigl[ (\xi\cdot \,^{o}e^I)\frac{\,^{1}e_I}{\rho}\bigr] =0\, , 
\end{equation}
since $\xi$ is constant on $S_\infty$ and $\d \,^{o}e^I=0$. On the other hand, it is also
easy to show that
\begin{equation}
\int_{S_\D} Q_{\xi\mathrm{HB}} = 0\, .
\end{equation}
since
\begin{equation}
\int_{S_\D}(\xi\cdot e^I)\, e^J\w\o_{IJ} =  \int_{S_\D} cl^I (e^J\w\o_{IJ})=0\, ,
\end{equation}
due to the expressions for the tetrad (\ref{hor1}) and connection (\ref{hor2}) on the horizon.

The variation of $S_{\Po}$, calculated from the LHS expression in (\ref{Pontryagin}), is
\be
\de S_{\Po}=-2 \int_{\mathcal{M}} DF^{IJ}\w\de\o_{IJ}+
2\int_{\partial \mathcal{M}}F^{IJ}\w\de\o_{IJ}\, ,\label{var_Po1}
\ee
so it does not contribute to the equations of motion in the bulk, due to the Bianchi
identity $DF^{IJ}=0$, and additionally the surface integral in
(\ref{var_Po1}) should vanish for the variational principle to be well defined. 
We will show later that this is indeed the case for boundary conditions of interest to us, namely asymptotically 
flat spacetimes possibly with an isolated horizon.

The symplectic potential current and the corresponding Noether charge 2-form for the Pontryagin term $S_{\Po}$, calculated from the LHS 
expression in (\ref{Pontryagin}) is
\be
\theta_\Po (\de )=2 F^{IJ}\w\de\o_{IJ}\, ,\ \ \ \ Q_{\xi\Po} =2 (\xi\cdot\o_{IJ})F^{IJ}\, .
\ee
We will show that the integrals of the Noether charge 2-form $Q_{\xi\Po}$ over $S_\infty$ and $S_\D$ vanish.
For the first one we have
\be
\int_{S_\infty} Q_{\xi\Po} = 2\int_{S_\infty} (\xi\cdot\o_{IJ})F^{IJ}=0\, .
\ee
since $\o_{IJ}=O(\rho^{-2})$, $F^{IJ}=O(\rho^{-3})$ and the volume element goes as $\rho^2$. 

Since the pull-back of the connection on $S_\D$ is given by (\ref{pullback_crossection}), 
we obtain that the integral of the Noether 2-form over $S_\D$ is 
\be\label{nch_po_hor}
\int_{S_\D} Q_{\xi\Po} =  
-2ic \int_{S_\D} (l\cdot V) \mathcal{R}\,\eps \, ,
\ee
where we have used the form of the connection on the horizon given by (\ref{hor2}). Note that the above
expression for the Noether charge is not gauge invariant on the horizon, under the rotations $m\to e^{i\theta}m$,  
the one-form $V$ transforms as $V\to V-i\d\theta$. So, in order to make the corresponding
Noether charge well defined we have to partially fix the gauge, by imposing $l\cdot\d\theta =0$.
This restricts the remaining gauge freedom 
$m\to e^{i\tilde{\theta}}m$ to the functions $\tilde{\theta}$ of the form  $\nabla_a\tilde{\theta}\,\heq\,w m_a+
\bar{w}\bm_a$, where $w$ is arbitrary.

On the other hand, we can calculate the symplectic potential current and the Noether charge 2-form from the 
RHS of (\ref{Pontryagin}), and obtain $\tilde{\theta}_\Po (\de )=\theta_\Po (\de )-2\d (\o^{IJ}\w\de\o_{IJ})$
and, as we have seen in (\ref{noether_charge_change}), this produces a following change in the Noether charge 2-form
\be
\tilde{Q}_{\xi\Po}=Q_{\xi\Po}-2\o^{IJ}\w\La_\xi\o_{IJ}\, .
\ee
It is easy to see that the integrals of the last term in the above equation over $S_{\infty}$ and $S_{\D}$ vanish,
due to our boundary conditions, hence the Noether charges remain invariant.

Similarly, for the Euler term, from the variation of the LHS of (\ref{Euler}), we obtain
\be
\theta_\Eu (\de )=2\, {\star F}^{IJ}\w\de\o_{IJ}\, ,\ \ \ \ Q_{\xi\Eu} =2(\xi\cdot\o_{IJ})\,{\star F}^{IJ}\, .
\ee
Then, as in case of the Pontryagin term it is easy to see that 
\be
\int_{S_\infty}Q_{\xi\Eu} = 2\int_{S_\infty}(\xi\cdot\o_{IJ})\,{\star F}^{IJ}=0\, ,
\ee
due to the asymptotic behaviour of the fields.

At the horizon the situation is different since the contraction of the dual of the pull-back 
(\ref{pullback_crossection}) is given by
\be
2\, l_{[I}n_{J]}\, {\star F}^{IJ}\vert_{S_\D}=-\mathcal{R}\,\eps\, ,
\ee
the corresponding Noether charge is non vanishing
\be
\int_{S_\D}Q_{\xi\Eu} = 2c \int_{S_\D} (l\cdot\o )\, \mathcal{R}\,\eps =16\pi c\,\kappa_{(l)}\label{noether_euler_horizon}
\ee
since $l\cdot\o =\k_{(l)}$ is constant on the horizon and the remaining integral is a topological invariant.
This result is consistent with the expression for the entropy of the Euler term in \cite{jacobson}, obtained in the second order formalism for stationary black holes. Note that, contrary to the Pontryagin term, the Noether charge of the Euler term
on the horizon is gauge invariant.

Though the Noether charge of the Euler term over a WIH is non-zero, the corresponding contribution to the Hamiltonian energy is nonetheless vanishing. 
As we have seen previously, in Section~\ref{sec:4.1}, the variation of the energy at the horizon is 
\be
\de H^\xi_\D = \int_{S_\D}\de Q_\xi -\xi\cdot\theta (\de )\, ,
\ee
with $\xi =cl$. For the Euler term we obtain
\be 
\int_{S_\D} cl\cdot\theta_\Eu (\de ) = 2c\int_{S_\D} l\cdot (\eps\w\de\o )\,\mathcal{R}=16\pi c\,\de\kappa_{(l)}\, ,
\ee
since $l\cdot\eps\,\heq\, 0$ and $l\cdot\de\o =\de\kappa_{(l)}$.
We see that this term cancels the variation of (\ref{noether_euler_horizon}) in  the expression for the energy at the horizon. 

Similarly as for the Pontryagin term, the variation of the RHS of (\ref{Euler}), leads to a change in the symplectic potential current and the Noether
charge 2-form, but the Noether charges stay invariant.

Finally, we have seen in Section~\ref{sec:4.3} that the variation of the Neih-Yan term on shell is proportional to the
variation of the Holst term, so all the results for the Noether charge of the Holst term apply directly here.
Namely, for the Neih-Yan term, with its boundary term, given in (\ref{well_defined_NY}), we obtain that
its Noether charge 2-form is 
\be
Q_{\xi\mathrm{NYB}}=2\k\g Q_{\xi\mathrm{HB}}\, ,
\ee
so that its integrals over $S_\infty$ and $S_\D$ vanish as well.

Let us end this section with a remark. One should note that the Noether charges at infinity of all the topological terms vanish for asymptotically flat boundary conditions,
but this is not the case for locally asymptotically anti-de Sitter (AAdS) space-times. In \cite{rodrigo} and \cite{aros},  
AAdS asymptotic conditions are considered and the Noether charge at 
infinity of the Palatini action with negative cosmological constant term turns out to be divergent.  
In that case the Euler term is added in order to make the action well defined
and finite. With this modification, the non vanishing (infinite) Noether charge becomes finite for
the well defined action. This illustrates that, in several respects, asymptotic AdS and asymptotically flat gravity behave in qualitatively different manners.

\section{Discussion and remarks}
\label{sec:5}

Let us start by summarizing the main results that we have here presented.

\begin{enumerate}

\item
We discussed the impact of the topological terms and boundary terms needed to have a well defined variational principle 
for any well posed field theory, on the symplectic
structure and the conserved Hamiltonian and Noether charges of the theory. We
showed, in particular, that for generic theories, no boundary term can modify the symplectic structure.

\item In the case of first order gravity, we showed that the topological terms do not modify the
symplectic structure.  In the case of the Holst term (that is {\it not} topological), there is a particular 
instance in which it could modify the symplectic structure.
Thus, the Hamiltonian structure of the theory remains
unaffected by the introduction of boundary and topological terms. In particular, all
Hamiltonian conserved quantities, that are generators of asymptotic symmetries, remain unaffected by such terms. 
We have also shown that for our boundary conditions the contribution from the Holst term to the Hamiltonian charges is always trivial. 
It is important to note that this simple result proves incorrect several
assertions that have repeatedly appeared in the literature.

\item We have shown that even when the Hamiltonian conserved charges remain
insensitive to the addition of boundary and topological terms, the
corresponding Noetherian charges {\it do} depend on such choices. This has as
a consequence that the identification of Noether charges with, say, energy
depends on the details of the boundary terms one has added. For instance, if
one focuses on the asymptotic region, then it is only for the well defined
Palatini with boundary action (of \cite{aes}) that the Noether charge coincides with
the Hamiltonian (ADM) energy. Any other choice, including Palatini without a boundary term,
would yield a different conserved quantity. Furthermore,
if one only had an internal boundary (and no asymptotic region), several
possibilities for the action are consistent (compare \cite{afk} and \cite{crv1}), 
and the relation between energy and Noether charge depends on such choices. We have
also made some comments regarding the relation between our analysis and others based on Noether
charges for stationary spacetimes \cite{iw}.

\end{enumerate}

In this manuscript, our focus was on first order gravity, but our analysis can be taken over to more general diffeomorphism invariant theories. Our results indicate that there
is an interesting interplay between symmetries and conserved quantities that
depends on the formalism used; Hamiltonian and Noether charges that have very different interpretations
within the theory, in general do not coincide. As we have seen, for the boundary conditions we considered most of the Noether charges associated to topological terms vanished --
while the Noether currents were non-vanishing--, but for generic
boundary conditions this might not be the case (such as in AADS asymptotics, for instance), indicating that generically these two sets of charges do not coincide. A deeper 
understanding of this issue is certainly called for.

Our analysis was done using the covariant Hamiltonian formalism, that has
proved to be economical and powerful to unravel the Hamiltonian structure of
classical gauge field theories.
It should be interesting to see whether a parallel analysis, using a 3+1 decomposition
of spacetime and taking special care on the effects of boundaries, yields similar results.
This work is in progress and will be reported elsewhere.

\section*{Acknowledgements}

We would like to thank N. Bodendorfer, S. Deser, T. Jacobson and R. Olea for comments. We would also like to thank an anonymous referee for comments that helped to improve the manuscript.
This work was in part supported by CONACyT 0177840, DGAPA-UNAM IN100212, 
and NSF PHY 1205388 grants, the Eberly Research Funds of Penn State
and by CIC, UMSNH.


\begin{thebibliography}{99}

\bibitem{romano} J.~D.~Romano,
  ``Geometrodynamics versus connection dynamics (in the context of (2+1) and (3+1) gravity,''
  Gen.\ Rel.\ Grav.\  {\bf 25}, 759 (1993)
  [gr-qc/9303032].

\bibitem{peldan} P.~Peldan,
  ``Actions for gravity, with generalizations: A Review,''
  Class.\ Quant.\ Grav.\  {\bf 11}, 1087 (1994)
  [gr-qc/9305011].

\bibitem{palatini} A. Palatini, ``Deduzione invariativa delle equazioni gravitazionali dal principio di Hamilton," 
Rend. Circ. Mat. Palermo {\bf 43}, 203 (1919).

\bibitem{adm} R.~L.~Arnowitt, S.~Deser and C.~W.~Misner,
  ``The Dynamics of general relativity,''
  Gen.\ Rel.\ Grav.\  {\bf 40}, 1997 (2008)
  [gr-qc/0405109].

\bibitem{utiyama} R.~Utiyama,
  ``Invariant theoretical interpretation of interaction,''
  Phys.\ Rev.\  {\bf 101}, 1597 (1956).

\bibitem{kibble} T.~W.~B.~Kibble,
  ``Lorentz invariance and the gravitational field,''
  J.\ Math.\ Phys.\  {\bf 2}, 212 (1961).

\bibitem{sciama} D.~W.~Sciama,
  ``The Physical structure of general relativity,''
  Rev.\ Mod.\ Phys.\  {\bf 36}, 463 (1964)
  [Erratum-ibid.\  {\bf 36}, 1103 (1964)].

\bibitem{di} S.~Deser and C.~J.~Isham,
  ``Canonical Vierbein Form of General Relativity,''
  Phys.\ Rev.\ D {\bf 14}, 2505 (1976).


\bibitem{holst} S.~Holst,
  ``Barbero's Hamiltonian derived from a generalized Hilbert-Palatini action,''
  Phys.\ Rev.\ D {\bf 53}, 5966 (1996)
  [gr-qc/9511026].

\bibitem{barros} N.~Barros e Sa,
  ``Hamiltonian analysis of general relativity with the Immirzi parameter,''
  Int.\ J.\ Mod.\ Phys.\ D {\bf 10}, 261 (2001)
  [gr-qc/0006013].

\bibitem{topo} S.~Deser, M.~J.~Duff and C.~J.~Isham,
  ``Gravitationally Induced Cp Effects,''
  Phys.\ Lett.\ B {\bf 93}, 419 (1980);
  A.~Ashtekar, A.~P.~Balachandran and S.~Jo,
  ``The {CP} Problem in Quantum Gravity,''  Int.\ J.\ Mod.\ Phys.\ A {\bf 4}, 1493 (1989);
   M.~Montesinos, ``Selfdual gravity with topological terms,''
  Class.\ Quant.\ Grav.\  {\bf 18}, 1847 (2001)
  [gr-qc/0104068].
  
\bibitem{aes} A.~Ashtekar, J.~Engle and D.~Sloan,
``Asymptotics and Hamiltonians in a First order formalism'',
Class.\ Quant.\ Grav.\ {\bf 25}, 095020 (2008).
[arXiv:0802.2527 [gr-qc]]
  
\bibitem{afk} A.~Ashtekar, S.~Fairhurst and B.~Krishnan,
``Isolated horizons: Hamiltonian evolution and the first law'',
Phys.\ Rev.\ {\bf D62}, 104025 (2000). 
[arXiv:0005083 [gr-qc]]  

\bibitem{apv} A.~Ashtekar, T.~Pawlowski and C.~Van Den
Broeck,
``Mechanics of higher-dimensional black holes in
asymptotically anti-de Sitter space-times'',
Class.\ Quant.\ Grav.\ {\bf 24}, 625 (2007). 
[arXiv: 0611049 [gr-qc]]

\bibitem{cwe} A.~Corichi and E.~Wilson-Ewing,
  ``Surface terms, Asymptotics and Thermodynamics of the Holst Action,''
  Class.\ Quant.\ Grav.\  {\bf 27}, 205015 (2010)
  [arXiv:1005.3298 [gr-qc]].
  
\bibitem{abr} A.~Ashtekar, L.~Bombeli and O.~Reula, 
``The covariant phase space of asymptotically flat gravitational fields'',
\textit{Analysis, Geometry and Mechanics: 200 Years after Lagrange}, eds. M.~Francaviglia and
D.~Holm, North-Holland, Amsterdam (1991).

\bibitem{witten} C.~Crnkovic and E.~Witten,
  ``Covariant Description of Canonical Formalism in Geometrical Theories,''
  \textit{Three hundred years of gravitation}, eds. Hawking, S.W. and  Israel, W., Cambridge U. Press (1987). 

\bibitem{wald-lee} J.~Lee and R.~M.~Wald,
  ``Local symmetries and constraints,''
  J.\ Math.\ Phys.\  {\bf 31}, 725 (1990);
R.~M.~Wald and A.~Zoupas,
  ``A General definition of 'conserved quantities' in general relativity and other theories of gravity,''
  Phys.\ Rev.\ D {\bf 61}, 084027 (2000)
  [gr-qc/9911095].
  
\bibitem{iw} V.~Iyer and R.~M.~Wald,
``Some properties of Noether charge and a proposal for dynamical
black hole entropy'',
Phys.\ Rev.\ {\bf D50}, 846 (1994).
[arXiv:9403028 [gr-qc]]

\bibitem{barbero1} J.~F.~Barbero G., J.~Prieto and E.~J.~S.~Villaseñor,
``Hamiltonian treatment of linear field theories in the presence of boundaries: a geometric approach'',
Class.\ Quant.\ Grav.\ {\bf 31}, 045021 (2014). [arXiv:1306.5854 [math-ph]]

\bibitem{crv1} A.~Corichi. I.~Rubalcava-Garc\'ia and T.~Vuka\v sinac, 
``Actions, topological terms and boundaries in first-order gravity: A review'',
Int.\ J.\ Mod.\ Phys.\ D {\bf 25}, 1630011 (2016). 


\bibitem{obukhov} Y.~N.~Obukhov,
``The Palatini principle for manifold with boundary'',
Class.\ Quantum\ Grav.\ {\bf 4}, 1085 (1987).

\bibitem{bn} N.~Bodendorfer and Y.~Neiman,
``Imaginary action, spinfoam asymptotics and the 'transplanckian' regime of loop quantum gravity''.
Class.\ Quantum\ Grav. {\bf 30}, 195018 (2013). [arXiv:1303.4752 [gr-qc]]


\bibitem{barbero} J.~F.~Barbero G.,
  ``Real Ashtekar variables for Lorentzian signature space times,''
  Phys.\ Rev.\ D {\bf 51}, 5507 (1995)
  [gr-qc/9410014].
  
\bibitem{nieh-yan} H.~T.~Nieh and M.~L.~Yan, 
``An identity in Riemann–Cartan geometry'', 
J.\ Math.\ Phys.\ {\bf 23}, 373 (1982).

\bibitem{nieh} H.~T.~Nieh, 
``A Torsional Topological Invariant'',
Int.\ J.\ Mod.\ Phys.\ {\bf A22}, 5237 (2007).   [arXiv:1309.0915 [gr-qc]]

\bibitem{W} R.~M.~Wald, ``General Relativity'', The University of Chicago Press, Chicago (1994).



\bibitem{chandra} S.~Chandrasekhar,
``The mathematical theory of black holes'',
Clanderon Press, Oxford (1992).

\bibitem{cg} A.~Chatterjee and A.~Ghosh,
``Laws of black hole mechanics from Holst action'',
Phys.\ Rev.\ {\bf D80}, 064036 (2009). 
[arXiv:0812.2121 [gr-qc]]


\bibitem{liko_booth} T.~Liko and I.~Booth,
``Isolated horizons in higher-dimensional Einstein-Gauss-Bonnet gravity'',
Class.\ Quant.\ Grav.\ {\bf 24}, 3769 (2007). [arXiv:0705.1371 [gr-qc]]

\bibitem{merced-holst} L.~Liu, M.~Montesinos and A.~Perez,
  ``A Topological limit of gravity admitting an SU(2) connection formulation,''
  Phys.\ Rev.\ D {\bf 81}, 064033 (2010)
  [arXiv:0906.4524 [gr-qc]].

\bibitem{liko_2} T.~Liko,
``Topological deformation of isolated horizons'',
Phys.\ Rev.\ {\bf D77}, 064004 (2008). [arXiv:0705.1518 [gr-qc]]

\bibitem{m:m} M.~Mondragon and M.~Montesinos,
  ``Covariant canonical formalism for four-dimensional BF theory,''
  J.\ Math.\ Phys.\  {\bf 47}, 022301 (2006)
  [gr-qc/0402041].

\bibitem{perez_2}
D.~J.~Rezende and A.~Perez,
  ``The Theta parameter in loop quantum gravity: Effects on quantum geometry and black hole entropy,''
  Phys.\ Rev.\ D {\bf 78}, 084025 (2008)
  [arXiv:0711.3107 [gr-qc]].
  
\bibitem{abl} A.~Ashtekar, C.~Beetle and J.~Lewandowski, 
``Mechanics of rotating isolated horizons''
Phys.\ Rev.\ {\bf D64}, 044016 (2001).
[arXiv:0103026 [gr-qc]]

\bibitem{pons} J.~M.~Pons,
``Boundary conditions from boundary terms, Noether charges and the trace K lagrangian in general relativity'',
Gen.\ Rel.\ Grav.\ {\bf 35}, 147 (2003).  [arXiv:0105032 [gr-qc]]


\bibitem{jacobson}  T.~Jacobson and R.~C.~Myers, 
``Black hole entropy and higher curvature interactions'',
Phys.\ Rev.\ Lett.\ {\bf 70}, 3684 (1993). 

\bibitem{rodrigo} R.~Aros, M.~Contreras, R.~Olea, R.~Troncoso and J.~Zanelli,
``Conserved charges for gravity with locally AdS asymptotics'',
Phys.\ Rev.\ Lett.\ {\bf 84}, 1647 (2000).  [arXiv:9909015 [gr-qc]]

\bibitem{aros} R.~Aros, 
``Boundary conditions in first order gravity: Hamiltonian and Ensemble'',
Phys.\ Rev.\  {\bf D73}, 024004 (2006).  [arXiv:0507091 [gr-qc]]













 

\end{thebibliography}
\end{document}